\documentclass[a4paper,12pt]{article}

\usepackage{amssymb}
\usepackage{amsmath}
\usepackage{graphicx}

\newtheorem{Prop}{Proposition}
\newtheorem{Def}[Prop]{Definition}
\newtheorem{Teo}[Prop]{Theorem}
\newtheorem{Lem}[Prop]{Lemma}

\newtheorem{Oss}[Prop]{Remark}
\newenvironment{Dim}
{\par\medskip\upshape\noindent\textbf{Proof}.}
{\hfill$\Box$\bigskip}

\newcommand{\D}{\mathrm{d}}
\newcommand{\BE}{\begin{equation}}
\newcommand{\EE}{\end{equation}}

\newcommand{\T}[1]{\mathbf{#1}}

\title{Complex variables for separation of Hamilton-Jacobi equation on real pseudo-Riemannian manifolds}
\author{Luca Degiovanni, Giovanni Rastelli}
\date{}

\begin{document}
\maketitle
\abstract{In this paper the geometric theory of separation of variables for time-independent Hamilton-Jacobi equation is extended to include the case of complex eigenvalues of a Killing tensor on pseudo-Riemannian manifolds. This task is performed without to complexify the manifold but just considering complex-valued functions on it. The simple formalism introduced allows to extend in a very natural way the classical results on separation of variables (including Levi-Civita criterion and St\"ackel-Eisenhart theory) to the complex case. Orthogonal variables only are considered.}

\section{Introduction}
The geometric theory of separation of variables for Hamilton-Jacobi equation on (pseudo)-Riemannian manifolds relates orthogonal separable coordinates with eigenvalues and eigenvectors of suitable Killing tensors.

If $Q$ is a real $n$-dimensional (pseudo)-Riemannian manifold with metric tensor $\T{g}$ and real coordinates $(q^\mu)$, then any symmetric contravariant two-tensor $\T{T}$ defines on the cotangent bundle $T^\ast Q$ the function $T=\frac{1}{2}T^{\mu\nu} p_\mu p_\nu$ where $p_\mu$ is the momentum conjugate to the coordinate $q^\mu$. Two tensors $\T{T}_1$ and $\T{T}_2$ are said to be in \emph{involution} or to \emph{commute in Schouten sense} if the corresponding functions are in involution with respect to the standard Poisson bracket on $T^\ast Q$. A tensor $\T{K}$ in involution with the metric is called a \emph{Killing tensor}. Killing $n$-tensors play a crucial role in Riemmanian geometry because they encode the symmetries of the metric, they can equivalently be defined by the equation $\nabla_{(\sigma} K_{\mu,\cdots,\nu)}=0$.

In the study of separation theory, it is usual to consider a single Killing two-tensor with pointwise distinct real eigenvalues and normal independent eigenvectors (called a \emph{characteristic Killing tensor}) or a family of $n$  independent Killing tensors (including the metric) that commute algebraically and in Schouten sense (called a \emph{Killing-St\"ackel algebra}). The main theorem \cite{E, KM1, KM2, Ben} that geometrically characterize the separability of the Hamilton--Jacobi equation states:
\begin{Teo}
The Hamilton--Jacobi equation associated to a natural geodesic Hamiltonian $H=\frac{1}{2}g^{\mu\nu}p_\mu p_\nu$ is additively separable in the orthogonal coordinates  $(q^\mu)$ if and only if a characteristic Killing tensor $\T{K}$ or, equivalently, a Killing-St\"ackel algebra exists, admitting the 1-forms $\D q^\mu$ as eigenforms. Being the coordinates orthogonal this is equivalent to ask that the vector fields $\frac{\partial}{\partial q^\mu}$ are eigenvectors. The coordinate hypersurfaces coincide with the foliation orthogonal to these eigenvectors. 
\end{Teo}
The two approaches, based respectively on characteristic Killing tensors or on Killing-S\"ackel algebras, are interchangeable, and we will use one or the other according to which is more suitable to the situation.

In pseudo-Riemannian manifolds (as the Minkowski space) real Killing tensors can have complex eigenvalues and eigenvectors. In previous papers the points of the manifold where this happens were considered as singular points, where no real separation coordinates were definable \cite{Ben}. Separable complex coordinates were introduced in~\cite{KM1,KM2} and subsequent papers on a complex Riemannian manifold. In this paper we introduce complex variables in a different way, avoiding the complexification of the original pseudo-Riemannian manifold. The complete integrals of the Hamilton-Jacobi equation found in this way turn out to be real-valued even if the variables used to determine them are complex-valued. This allow us to extend the definition of characteristic Killing tensor to pseudo-Riemannian manifolds without asking the eigenvalues to be real and do not require extension on complex manifold of the Hamilton-Jacobi theory.

\section{Complex variables from a Killing tensor}
Let $\T{K}$ be a Killing tensor on the real $n$-dimensional pseudo-Riemannian manifold $Q$ (with real coordinates ($q^\mu$) and metric tensor $\T{g}$), such that $n$ pointwise distinct functions $\lambda^\mu$ and $n$ functionally independent functions $z^\mu$ are defined in a open subset of $Q$ and satisfy (with no summation over $\mu$) the equations:
\BE\label{autovet}
\T{K}\D z^\mu =\lambda^\mu \T{g}\D z^\mu\,.
\EE
In other terms, $\T{K}$ admits $n$ independent exact eigenforms  $\D z^\mu$, corresponding to the $n$ pointwise distinct eigenvalues $\lambda^\mu$. We call this kind of Killing tensor a \emph{characteristic Killing tensor}, dropping the request that the functions $\lambda^\mu$ (and as a consequence the functions $z^\mu$) are real. Indeed, being the metric $\T{g}$ not positively defined, the functions $\lambda^\mu$ and $z^\mu$ can assume complex values. Because $\T{K}$ and $\T{g}$ are both symmetric with real coefficients,  if $z^\mu$ satisfies the relation~(\ref{autovet}) also its complex conjugate $\bar{z}^\mu$ satisfies the same relation. Hence, the functions $z^\mu$ with a non-vanishing imaginary part are organized in pairs and the $n$ complex-valued functions $z^\mu$ define, through their real and imaginary parts, $n$ real-valued coordinates.

It is important to observe that also when $\T{K}$ has complex eigenvalues it is not necessary to complexify  the manifold $Q$: the functions $z^\mu$ are defined by using the real coordinates $(q^\mu)$, even if they have complex values. In this paper all the functions $z^\mu$ are sometimes treated as independent, without taking in account the conjugation relation and using distinct indices for the function $z^\mu$ and the function $\bar{z}^\mu$, also indicate by $z^{\bar{\mu}}$. 

This point of view is preferable to the use of the real and imaginary part of  the functions $z^\mu$ for at least two reason: because the simpler notation allow to handle the complex case in a way analogous to the ordinary real case and, more substantially, because the pairs of conjugate variables can change from a point on $Q$ to another. On a three dimensional manifold, for example, in each point at least one of the three function $z^\mu$ is real-valued but any of them can assume complex values in some subset of $Q$. The use of different indices for a function and its complex conjugate allows to not specify which variables are real and which belong to a conjugate pair. In any case, in a neighborhood of any point of $Q$ it is possible to split the functions $z^\mu$ in a subset of real functions and in a subset of conjugate pairs. The real ones will be denoted by latin indices: $z^a$, while the conjugate pairs will be denoted by greek ones: $z^\alpha$ and $z^{\bar{\alpha}}=\bar{z}^\alpha$.

When the functions $z^\mu$ assume complex values the definition of the derivation $\frac{\partial}{\partial z^\mu}$  is no more obvious and some extra care is needed to use the $z^\mu$ as a system of variables on the real manifold $Q$. Anyway, because of the conjugation relation, one must be careful to consider the $z^\mu$ as coordinates. We will see that this abuse of terminology can lead to some incongruence, hence, we will call the functions $z^\mu$ simply ``variables''.

Considering the Jacobian matrix $\left(\frac{\partial z^\mu}{\partial q^\nu}\right)$ and its inverse $\T{J}$ one can define the, eventually complex-valued, vector fields
\BE\label{zder}
Z_\mu=\frac{\partial}{\partial z^\mu}=J^\nu_\mu\frac{\partial}{\partial q^\nu}\,.
\EE
\begin{Lem}\label{Lem_der}
The vector fields~(\ref{zder}) satisfy the following properties:
\begin{enumerate}
\item $Z_\mu(q^\nu)=J^\nu_\mu$;
\item $Z_\mu(z^\nu)=\delta^\nu_\mu$ and if $\bar{z}^\mu\neq z^\mu$ then $Z_\mu(\bar{z}^\mu)=0$;
\item if $z^{\bar{\mu}}=\bar{z}^\mu$ then $Z_{\bar{\mu}}=\overline{Z}_\mu$;
\item $[Z_\mu,Z_\nu]=0$;
\item $\overline{Z}_\mu(\overline{W})=\overline{Z_\mu(W)}$.
\end{enumerate}
\begin{Dim}
The first identity follows directly from the definition~(\ref{zder}), applying the vector field to the old coordinates $(q^\nu)$.
The second property holds because $\T{J}$ is the inverse matrix of $\left(\frac{\partial z^\mu}{\partial q^\sigma}\right)$.
The third is obtained from
$$
J^\nu_{\bar{\mu}}\frac{\partial z^{\bar{\mu}}}{\partial q^\sigma}=\delta_\sigma^\nu=\bar{J}^\nu_\mu\overline{\frac{\partial z^\mu}{\partial q^\sigma}}\,,
$$
because $\frac{\partial z^{\bar{\mu}}}{\partial q^\sigma}=\overline{\frac{\partial z^\mu}{\partial q^\sigma}}$ one has $J^\nu_{\bar{\mu}}=\bar{J}^\nu_\mu$ and being $Z_{\bar{\mu}}=J^\nu_{\bar{\mu}}\frac{\partial}{\partial q^\nu}$ the thesis follows.
To show that $Z_\mu$ and $Z_\nu$ commute, a longer calculation is needed: from the definition of commutator one has
$$
[Z_\mu,Z_\nu]=\left(J^\sigma_\mu\frac{\partial}{\partial q^\sigma} J^\tau_\nu-J^\sigma_\nu\frac{\partial}{\partial q^\sigma} J^\tau_\mu\right)\frac{\partial}{\partial q^\tau}
$$
by differentiating with respect to $q^\sigma$ the identity
$$
J^\tau_\rho\frac{\partial z^\lambda}{\partial q^\tau}=\delta^\lambda_\rho\,,
$$
and setting in one case $\rho=\mu$ and in the other $\rho=\nu$, it follows that
$$
[Z_\mu,Z_\nu](z^\lambda)=J^\sigma_\mu J^\tau_\nu\frac{\partial^2 z^\lambda}{\partial q^\sigma\partial q^\tau}-J^\sigma_\nu J^\tau_\mu\frac{\partial^2 z^\lambda}{\partial q^\sigma\partial q^\tau}=0\,.
$$
Finally, the last identity is proved observing that
$$
\bar{Z}_\mu(\bar{W})=\bar{J}^\sigma_\mu\frac{\partial\bar{W}}{\partial q^\sigma}=\overline{J^\sigma_\mu\frac{\partial W}{\partial q^\sigma}}\,.
$$
\end{Dim}
\end{Lem}
The vector field $Z_\mu$ allows to differentiate a (possibly complex-valued) function $f$ on $Q$ with respect to the variable $z^\mu$:
$$
\frac{\partial f}{\partial z^\mu}=Z_\mu(f)=J^\nu_\mu\frac{\partial f}{\partial q^\nu}=
\frac{\partial q^\nu}{\partial z^\mu}\frac{\partial f}{\partial q^\nu}\,.
$$

For the variables $z^\alpha$ and $z^{\bar{\alpha}}$ belonging to a conjugate pair the vector fields associated to their real and imaginary parts can be introduced.
\begin{Lem}
If $z^\alpha=x^\alpha(q^\mu)+iy^\alpha(q^\mu)$ then the real-valued vector fields
\begin{eqnarray*}
\frac{\partial}{\partial x^\alpha}&=&\frac{\partial}{\partial z^\alpha}+\frac{\partial}{\partial z^{\bar{\alpha}}}\\
\frac{\partial}{\partial y^\alpha}&=&i\left(\frac{\partial}{\partial z^\alpha}-\frac{\partial}{\partial z^{\bar{\alpha}}}\right)
\end{eqnarray*}
can be defined and they satisfies the relations $\frac{\partial}{\partial x^\alpha} x^\alpha=\frac{\partial}{\partial y^\alpha} y^\alpha=1$, $\frac{\partial}{\partial x^\alpha} y^\alpha=\frac{\partial}{\partial y^\alpha} x^\alpha=0$. Hence
\BE\label{zder_comp}
\begin{array}{rcl}
\dfrac{\partial}{\partial z^\alpha}&=&\dfrac{1}{2}\dfrac{\partial}{\partial x^\alpha} - \dfrac{i}{2}\dfrac{\partial}{\partial y^\alpha}\,,\\[10pt]
\dfrac{\partial}{\partial z^{\bar{\alpha}}}&=&\dfrac{1}{2}\dfrac{\partial}{\partial x^\alpha} + \dfrac{i}{2}\dfrac{\partial}{\partial y^\alpha}\,.
\end{array}
\EE
\begin{Dim}
Being $z^\alpha\neq z^{\bar{\alpha}}$ the two vector fields $Z_\alpha$ and $Z_{\bar{\alpha}}$ are different, thus both the vector fields $\frac{\partial}{\partial x^\alpha}$ and $\frac{\partial}{\partial y^\alpha}$ are well-defined. Applying these two vector fields to the functions $x^\alpha=\frac{1}{2}(z^\alpha+z^{\bar{\alpha}})$ and  $y^\alpha=\frac{i}{2}(z^{\bar{\alpha}}-z^\alpha)$ the desired relations are found. The expression of $Z_\alpha$ and $Z_{\bar{\alpha}}$ in terms of the vector fields $\frac{\partial}{\partial x^\alpha}$ and $\frac{\partial}{\partial y^\alpha}$ is a trivial consequence of the definition.
\end{Dim}
\end{Lem}
The expressions~(\ref{zder_comp}) are sometimes called \emph{Wirtinger formal derivatives}. Given $z^\alpha$ and $z^\beta$ both with non-vanishing imaginary part and $\beta$ different from $\alpha$ and $\bar{\alpha}$ then, because $x^\beta$ and $y^\beta$ depend only on $z^\beta$ and $z^{\bar{\beta}}$, one has $\frac{\partial}{\partial x^\alpha}x^\beta=\frac{\partial}{\partial y^\alpha}y^\beta=0$ \cite{Hor}.

The formulas~(\ref{zder_comp}) are widely used in complex analysis. As instance if $W=A(x^\mu,y^\mu)+iB(x^\mu,y^\mu)$ the equation $\frac{\partial}{\partial \bar{z}^\mu}W=0$ is equivalent to Cauchy-Riemann equations. As an example of a misleading due to considering $z^\mu$ and $\bar{z}^\mu$ as coordinates, let us consider the Hartogs Theorem (\cite{Hor}, Theorem 2.2.8):
\begin{Teo}\label{olo}
Let $f(z^1,\dots ,z^n)$ a complex function defined on a open subset of $\mathbb{C}$, then $f$ is holomorphic if and only if it is separately holomorphic in each variables, i.e. if and only if
\BE
\displaystyle \frac{\partial f}{\partial \bar z^\sigma}=0, \qquad \sigma =1, \dots, n.
\EE
\end{Teo}
It is obvious that $W=W_1(z^\mu)+W_2(z^\nu)$ is separately olomorphic, but if $z^\nu=\bar{z}^\mu$ then does not follows that $W$ is olomorphic.

The efficiency of the introduced formalism becomes particularly evident in the definition of the  momenta $P_\mu$ conjugate to the variables $z^\mu$ through the formulas
\BE\label{trasf_can}
\begin{array}{l}
z^\mu=z^\mu(q^\nu) \\[10pt]
p_\mu=\dfrac{\partial z^\nu}{\partial q^\mu}P_\nu
\end{array}
\EE
formally analogous to the canonical transformation associated to a change of coordinates in the manifold $Q$. If $z^\mu$ is a complex variable also $P_\mu$ is complex. Thanks to Lemma~\ref{Lem_der} one can define the conjugate momentum $P_\nu$ as:
\BE\label{trasf_inv}
P_\nu=J^\mu_\nu p_\mu=\frac{\partial q^\mu}{\partial z^\nu}p_\mu\,.
\EE

The canonical Poisson bracket on $T^\ast Q$ can be extended, by bilinearity, to complex-valued functions and the transformation (\ref{trasf_can}) turns out to be canonical and to behave well under complex conjugation.

The variables $(P_\mu,z^\mu)$ are widely used in the literature on Hamiltonian systems  (e.g. see \cite{Puc}). The main properties of these variables are recalled for sake of clearness.

\begin{Lem}
The relations \mbox{$\{P_\mu,z^\nu\}=\delta^\nu_\mu$}, \mbox{$\{z^\mu,z^\nu\}=0$} and \mbox{$\{P_\mu,P_\nu\}=0$} hold. Therefore the previously defined variables \mbox{$(P_\mu$, $z^\mu)$} are canonical.
\begin{Dim}
The relations involving just the functions $z^\mu$ are obvious, because they depends only on the $(q^\mu)$. The other two relations are proved considering that the elements of the matrix $\left(\frac{\partial q^\nu}{\partial z^\mu}\right)$ depend only on the $(q^\mu)$ and using the standard argument:
\begin{eqnarray*}
\{P_\mu,z^\nu\} &=& \{\frac{\partial q^\sigma}{\partial z^\mu}p_\sigma,z^\nu\} \\
&=& \frac{\partial q^\sigma}{\partial z^\mu}\{p_\sigma,z^\nu\}+\{\frac{\partial q^\sigma}{\partial z^\mu},z^\nu\}p_\sigma \\
&=& \frac{\partial q^\sigma}{\partial z^\mu}\frac{\partial z^\nu}{\partial q^\sigma}=\delta^\nu_\mu\\
\{P_\mu,P_\nu\} &=& \{\frac{\partial q^\sigma}{\partial z^\mu}p_\sigma,\frac{\partial q^\tau}{\partial z^\nu}p_\tau\} \\
&=& \frac{\partial q^\sigma}{\partial z^\mu}\frac{\partial q^\tau}{\partial z^\nu}\{p_\sigma,p_\tau\} +
\frac{\partial q^\sigma}{\partial z^\mu}\{p_\sigma,\frac{\partial q^\tau}{\partial z^\nu}\}p_\tau +\\
&&\frac{\partial q^\tau}{\partial z^\nu}\{\frac{\partial q^\sigma}{\partial z^\mu},p_\tau\}p_\sigma +
\{\frac{\partial q^\sigma}{\partial z^\mu},\frac{\partial q^\tau}{\partial z^\nu}\}p_\sigma p_\tau\\
&=& \frac{\partial q^\sigma}{\partial z^\mu}p_\tau\frac{\partial}{\partial q^\sigma}\frac{\partial q^\tau}{\partial z^\nu} -\frac{\partial q^\tau}{\partial z^\nu}p_\sigma\frac{\partial}{\partial q^\tau}\frac{\partial q^\sigma}{\partial z^\mu} \\
&=& \frac{\partial q^\sigma}{\partial z^\mu}p_\tau\frac{\partial}{\partial z^\nu}\delta^\tau_\sigma-\frac{\partial q^\tau}{\partial z^\nu}p_\sigma\frac{\partial}{\partial z^\mu}\delta^\sigma_\tau =0
\end{eqnarray*}
\end{Dim}
\end{Lem}
\begin{Lem}
If $z^{\bar{\mu}}=\bar{z}^\mu$ then $P_{\bar{\mu}}=\overline{P_\mu}$.
\begin{Dim}
This property is immediately deduced taking the complex conjugate of the formula (\ref{trasf_inv}) and recalling that $\frac{\partial q^\nu}{\partial \bar{z}^\mu}=\overline{\frac{\partial q^\nu}{\partial z^\mu}}$ and that the $p_\mu$ are real-valued functions.
\end{Dim}
\end{Lem}

The real and imaginary parts $x^\alpha$ and $y^\alpha$ of $z^\alpha$ can be used to define the correspondent real-valued conjugate momenta.
\begin{Lem}
If $z^\alpha=x^\alpha+iy^\alpha$ has a non-vanishing imaginary part then, set $z^\alpha=z$, $P_\alpha=P_z$, $x^\alpha=x$ and $y^\alpha=y$, the complex momentum $P_z$ can be written as a function of the real momenta $P_x$ and $P_y$, conjugate to the coordinates $x$ and $y$, through  the formula
\BE\label{p_Z comp}
P_z=\frac{1}{2}(P_x-i P_y)
\EE
\begin{Dim}
If all the variables $z^\mu$ with $\mu$ different from $\alpha$ and $\bar{\alpha}$ are left unchanged, the transformation
\begin{eqnarray*}
z=x+iy \quad \bar{z}=x-iy \\
P_x=\frac{\partial z}{\partial x}P_z+\frac{\partial \bar{z}}{\partial x}P_{\bar{z}}  &=& P_z+P_{\bar{z}} \\
P_y=\frac{\partial z}{\partial y}P_z+\frac{\partial \bar{z}}{\partial y}P_{\bar{z}} &=& i(P_z-P_{\bar{z}})
\end{eqnarray*}
is canonical and defines the real conjugate momenta $P_x$ and $P_y$. The formula (\ref{p_Z comp}) can be obtained, by solving the previous equations with respect to $P_z$.
\end{Dim}
\end{Lem}
In order to write in the variables $(z^\mu$, $P_\mu)$ the Poisson bracket and the Hamilton equations, the vector fields 
$$
\frac{\partial}{\partial P_\mu}=\frac{\partial z^\mu}{\partial q^\nu}\frac{\partial}{\partial p_\nu}
$$
have to be introduced. As a consequence $\frac{\partial}{\partial p_\nu}=J^\nu_\mu\frac{\partial}{\partial P_\mu}$. If $z^\alpha$ has non-vanishing imaginary part then by setting $z^\alpha=z=x+iy$ and being $P_\alpha=P_z=\frac{1}{2}(P_x-iP_y)$ its conjugate momentum one can write
\begin{eqnarray*}
\frac{\partial}{\partial P_\alpha}&=&\frac{\partial}{\partial P_x}+i\,\frac{\partial}{\partial P_y}\\
\frac{\partial}{\partial P_{\bar{\alpha}}}&=&\frac{\partial}{\partial P_x}-i\,\frac{\partial}{\partial P_y}
\end{eqnarray*}
\begin{Lem}\label{lemmino}
The following properties hold:
\begin{enumerate}
\item $\left[\dfrac{\partial}{\partial P_\mu},\dfrac{\partial}{\partial P_\nu}\right]=0$;
\item $\left[\dfrac{\partial}{\partial z^\mu},\dfrac{\partial}{\partial P_\nu}\right]=0$;
\item if $z^{\bar{\mu}}=\bar{z}^\mu$ then $\dfrac{\partial}{\partial P_{\bar{\mu}}}=\overline{\dfrac{\partial}{\partial P_\mu}}$.
\end{enumerate}
\begin{Dim}
The first relation follows because $\frac{\partial z^\mu}{\partial q^\nu}$ does not depend on $p_\sigma$, while the third is a consequence of Lemma~\ref{Lem_der}. The second property is obtained by differentiating the identity
$$
\frac{\partial q^\tau}{z^\mu}\frac{\partial z^\nu}{\partial q^\tau}=\delta_\mu^\nu
$$
with respect to $q^\sigma$ and observing that $\frac{\partial q^\tau}{\partial q^\sigma}=0$ implies
$$
\frac{\partial q^\tau}{\partial z^\mu}\frac{\partial^2 z^\nu}{\partial q^\tau\partial q^\sigma}=0\,.
$$
Hence the thesis follows from
$$
\left[\frac{\partial}{\partial z^\mu},\frac{\partial}{\partial P_\nu}\right]=\frac{\partial q^\tau}{\partial z^\mu}\frac{\partial^2 z^\nu}{\partial q^\tau\partial q^\sigma}\frac{\partial}{\partial p_\sigma}\,.
$$
\end{Dim}
\end{Lem}
\begin{Prop}
The Poisson bracket of two, possibly complex-valued, functions $f(z^\mu,P_\mu)$ and $g(z^\mu,P_\mu)$ is  given by
$$
\{f,g\}=\frac{\partial f}{\partial P_\mu}\frac{\partial g}{\partial z^\mu}-
\frac{\partial f}{\partial z^\mu}\frac{\partial g}{\partial P_\mu}\,.
$$
Hence, the Hamilton equations for an Hamiltonian $H(z^\sigma,P_\sigma)$ are
$$
\left\{\begin{array}{rcl}
\dot{z}^\mu&=&\dfrac{\partial H}{\partial P_\mu} \\[10pt]
\dot{P}_\mu&=&-\dfrac{\partial H}{\partial z^\mu}
\end{array}
\right.
$$
\begin{Dim}
\begin{eqnarray*}
\{f,g\}&=&\frac{\partial f}{\partial p_\sigma}\frac{\partial g}{\partial q^\sigma}-\frac{\partial f}{\partial q^\sigma}\frac{\partial g}{\partial p_\sigma}\\
&=& \frac{\partial q^\sigma}{\partial z^\mu}\frac{\partial f}{\partial P_\mu}\frac{\partial z^\nu}{\partial q^\sigma}\frac{\partial g}{\partial z^\nu}
-\frac{\partial z^\nu}{\partial q^\sigma}\frac{\partial f}{\partial z^\nu}\frac{\partial q^\sigma}{\partial z^\mu}\frac{\partial g}{\partial P_\mu}\\
&=& \frac{\partial f}{\partial P_\mu}\frac{\partial g}{\partial z^\mu}
-\frac{\partial f}{\partial z^\mu}\frac{\partial g}{\partial P_\mu}
\end{eqnarray*}
\end{Dim}
\end{Prop}

\begin{Oss} Due to the conjugation relation between $z^\mu$ and $z^{\bar \mu}$ we have that the splitting of these complex Hamilton equations into real and imaginary parts yields two copies of the equivalent real Hamilton equations in real coordinates. This will happen again in the following and we denote the phenomenon as "`redundancy"'.
\end{Oss}

The next Proposition is fundamental  for separation of variables theory in complex variables:
\begin{Prop}
The metric $\T{g}$ and the characteristic Killing tensor $\T{K}$ are simultaneously diagonalized in the variables $(z^\mu,P_\mu)$. In other words, the two associated quadratic functions
$$
H=\frac{1}{2}g^{\mu\nu}p_\mu p_\nu\,, \quad K=\frac{1}{2}K^{\mu\nu}p_\mu p_\nu
$$
assumes the forms
$$
H=\frac{1}{2}{\textstyle \sum_\sigma}g^\sigma{P_\sigma}^2\,, \quad K=\frac{1}{2}{\textstyle \sum_\sigma}\lambda^\sigma g^\sigma{P_\sigma}^2\,.
$$
\begin{Dim}
By the transformation formula (\ref{trasf_can}) in the new variables one has
$$
H=\frac{1}{2}g^{\mu\nu}\frac{\partial z^\sigma}{\partial q^\mu}\frac{\partial z^\tau}{\partial q^\nu}P_\sigma P_\tau\,, \quad K=\frac{1}{2}K^{\mu\nu}\frac{\partial z^\sigma}{\partial q^\mu}\frac{\partial z^\tau}{\partial q^\nu}P_\sigma P_\tau
$$
but $\lambda^\sigma g^{\mu\nu}\partial_\mu z^\sigma\partial_\nu z^\tau=K^{\mu\nu}\partial_\mu z^\sigma\partial_\nu z^\tau=\lambda^\tau g^{\mu\nu}\partial_\mu z^\sigma\partial_\nu z^\tau$ and thus (being $\lambda^\sigma$ pointwise distinct functions) if $\sigma\ne\tau$ it follows $g^{\mu\nu}\partial_\mu z^\sigma\partial_\nu z^\tau=0$ and $K^{\mu\nu}\partial_\mu z^\sigma\partial_\nu z^\tau=\lambda^\sigma g^{\mu\nu}\partial_\mu z^\sigma\partial_\nu z^\tau=0$. The thesis is obtained by setting, for the non-vanishing coefficients,
$$
g^{\mu\nu}\frac{\partial z^\sigma}{\partial q^\mu}\frac{\partial z^\sigma}{\partial q^\nu}=g^\sigma\, \qquad K^{\mu\nu}\frac{\partial z^\sigma}{\partial q^\mu}\frac{\partial z^\sigma}{\partial q^\nu}=K^\sigma\,.
$$

\end{Dim}
\end{Prop}

Redundancy manifest itself here by the straightforward relations $\bar {g^\sigma}=g^{\bar \sigma}$ and $\bar {K^\sigma}=K^{\bar \sigma}$.
\begin{Oss}
Given a characteristic Killing tensor $K^{\mu\nu}$, or the associated Killing-St\"ackel algebra, the functions $z^\mu$ are given by integrating either the eigenforms or the eigenvectors of $K^{\mu\nu}$  or (as we will see) by the eigenvalues of a basis of the associated Killing-St\"ackel algebra through the fundamental functions $f_a^{bc}$ given in \cite{CR}, generalized to the complex case.
\end{Oss}

\section{The Hamilton--Jacobi equation and its solutions}
The Hamilton--Jacobi equation for  Hamiltonian $H$, written in complex variables, is obtained by setting
\BE\label{momenti_pos}
P_\mu=\frac{\partial W}{\partial z^\mu}\,.
\EE
in the expression of $H$. In this way a PDE with complex coefficients for a function $W$ is found and in general its solutions will be complex-valued. The consistence of the substitution~(\ref{momenti_pos}) implies that $P_{\bar{\mu}}=\overline{P}_\mu$ if $z_{\bar{\mu}}=\bar{z}_\mu$. This fact is achieved only by using a real-valued solution of the Hamilton--Jacobi equation. 

\begin{Prop}\label{reale}
Let $H(P_\mu,z^\mu)$ be a time-independent Hamiltonian written in the canonical complex variables $(P_\mu, z^\mu)$. A solution $W$ of the Hamilton--Jacobi equation associated to $H$:
\begin{equation}\label{HJ}
H(\frac{\partial W}{\partial z^\mu},z^\mu)=h
\end{equation}
has a constant imaginary part (and then an equivalent real solution exists) if an only if $\frac{\partial W}{\partial z^\mu}$ is complex conjugate to $\frac{\partial W}{\partial \bar{z}^\mu}$.
\begin{Dim}
From Lemma~\ref{Lem_der} it follows that (set $z^{\bar{\mu}}=\bar{z}^\mu$)
\begin{eqnarray*}
\frac{\partial W}{\partial z^\mu}=J^\nu_\mu \frac{\partial W}{\partial q^\nu} \\
\frac{\partial W}{\partial z^{\bar{\mu}}}=\bar{J}^\nu_\mu \frac{\partial W}{\partial q^\nu} 
\end{eqnarray*}
Hence, being the matrix $\T{J}$ invertible, the first equation is the complex conjugate of the second one if and only if
$$
\frac{\partial W}{\partial q^\nu}=\bar {\frac{\partial W}{\partial q^\nu}}
$$
that is if and only if the imaginary part of $W$ is constant (and is equivalent to zero).
\end{Dim}
\end{Prop}

The relation between the Hamilton--Jacobi equation with complex variables and the ordinary one can be clarified by using (when $z^\alpha=z=x+iy$ has a non-vanishing imaginary part) the non orthogonal coordinates given by their real and imaginary parts. Being
$$
\frac{\partial W}{\partial z^\alpha}=\frac{1}{2}\frac{\partial W}{\partial x}-\frac{i}{2}\frac{\partial W}{\partial y},
$$ 
by using the formula (\ref{p_Z comp}) one obtain the substitution employed in the ordinary Hamilton--Jacobi equation:
$$
P_x=\frac{\partial W}{\partial x}\,, \quad P_y=\frac{\partial W}{\partial y}\,.
$$

\begin{Def}\label{OLO}
A \emph{complete integral} of the equation~(\ref{HJ}) is a solution $W(z^\mu,c_\nu)$ of (\ref{HJ}), where $c_\nu$ are $n$ real-valued parameters, satisfying
$$
\det \left(\frac{\partial^2 W}{\partial z^\mu \partial c_\nu}\right)\neq0.
$$
A complete integral $W(z^\mu,c_\nu)$ is (additively) separated in the variables $(z^\mu)$ if
$$
W(z^\mu,c_\nu)=W_1(z^1,c_\nu)+\dots + W_n(z^n,c_\nu).
$$  
\end{Def}
The definition (\ref{zder}) of $\frac{\partial}{\partial z^\mu}$ implies
\BE
\det \left(\frac{\partial^2 W}{\partial z^\mu \partial c_\nu}\right)=\det\T{J}\,\det \left(\frac{\partial^2 W}{\partial q^\sigma \partial c_\nu}\right)
\EE
then, because $\T{J}$ is invertible, when $W$ is real the definition of complete integral in the complex case is equivalent to the usual one. It is important to notice that by rewriting the function $W$ in the coordinates $(q^\mu)$ one obtains different results if the variables $(z^\mu)$ are complex or real.
The introduction of the complex variables allows to find a complete real integral of the Hamilton--Jacobi equation and then to apply the classical Jacobi method even where the eigenvalues of $\T{K}$ are not real. 

\section{Separation of variables for Hamilton--Jacobi equation}
The Levi--Civita criterion for the complete additive separation of a Hamilton--Jacobi equation can be rewritten in our complex variables. Following Levi--Civita \cite{LC}, the Hamiltonians considered will be always non-degenerated, that is satisfying $\displaystyle \frac {\partial H}{\partial P_\mu}\neq 0$ for all the indices, but not necessarily natural. Moreover, it is also possible to consider a complex valued Hamiltonian.
\begin{Teo}
The Hamilton--Jacobi equation for Hamiltonian $H$ admits a complete integral separated in the variables $(z^\mu)$, if and only if the Hamiltonian $H$ satisfies the Levi--Civita equation
$$
\partial^\mu H \partial^\nu H \partial_{\mu\nu} H+
\partial_\mu H \partial_\nu H \partial^{\mu\nu} H-
\partial^\mu H \partial_\nu H \partial_\mu^\nu H-
\partial_\mu H \partial^\nu H \partial^\mu_\nu H=0
$$
where $\mu\neq\nu$, no summation over the repeated indices is understood and
$$
\partial_\mu=\frac{\partial}{z^\mu}=\frac{\partial q^\nu}{\partial z^\mu}\frac{\partial}{\partial q^\nu}\,,
\quad
\partial^\mu=\frac{\partial}{P_\mu}=\frac{\partial z^\mu}{\partial q^\nu}\frac{\partial}{\partial p_\nu}\,.
$$
\begin{Dim}
The proof is adapted from the Levi--Civita's one \cite{LC}. The Hamilton--Jacobi equation for the Hamiltonian $H(P_\mu,z^\mu)$ is $H(w_\mu,z^\mu)=h$ where $P_\mu=\frac{\partial W}{\partial z^\mu} =w_\mu(z^\nu)$. By differentiating this equation with respect to $z^\nu$ one obtains
$$
\frac{\partial H}{\partial z^\nu}+\frac{\partial H}{\partial P_\mu}\frac{\partial w_\mu}{\partial z^\nu}=0\,.
$$
The additive separation of variables is obtained if and only if $w_\mu$ depends only from $z^\mu$. If $\frac{\partial H}{\partial P_\mu}\neq0$ the complete separability is equivalent to the existence of a solution for each of the $n$ PDE system
\BE\label{sistema}
\frac{\partial w_\mu}{\partial z^\mu}=-\frac{\frac{\partial H}{\partial z^\mu}}{\frac{\partial H}{\partial P_\mu}}=R_\mu(w_\sigma,z^\sigma)\,,
\quad
\frac{\partial w_\mu}{\partial z^\nu}=0 \; (\nu\neq\mu)\,.
\EE
Each PDE system (\ref{sistema}) can be seen as a system for a function depending on $n$ independent complex variables, but the variables $z^\mu$ obtained from (\ref{autovet}) are not independent due to complex conjugation relations. Let us split the functions $z^\mu$ in real ones and conjugate pairs: $(z^\mu)=(z^a,z^\alpha,z^{\bar{\alpha}})$. The system (\ref{sistema}) can be rewritten using the Wirtinger formal derivatives (\ref{zder_comp}) and, depending on the value of $\mu$, one of the following systems is obtained
\begin{equation}\label{sis_reale}
\left\{\begin{array}{l}
\frac{\partial w_a}{\partial z^a}=R_a \\[5pt]
\frac{\partial w_a}{\partial z^b}=0 \\[5pt]
\frac{\partial w_a}{\partial x^\alpha}=0 \\[5pt]
\frac{\partial w_a}{\partial y^\alpha}=0
\end{array}\right.
\quad
\left\{\begin{array}{l}
\frac{\partial w_\alpha}{\partial x^\alpha}=R_\alpha \\[5pt]
\frac{\partial w_\alpha}{\partial y^\alpha}=iR_\alpha \\[5pt]
\frac{\partial w_\alpha}{\partial x^\beta}=0 \\[5pt]
\frac{\partial w_\alpha}{\partial y^\beta}=0 \\[5pt]
\frac{\partial w_\alpha}{\partial z^a}=0
\end{array}\right.
\quad
\left\{\begin{array}{l}
\frac{\partial w_{\bar{\alpha}}}{\partial x^\alpha}=R_{\bar{\alpha}} \\[5pt]
\frac{\partial w_{\bar{\alpha}}}{\partial y^\alpha}=-iR_{\bar{\alpha}} \\[5pt]
\frac{\partial w_{\bar{\alpha}}}{\partial x^\beta}=0 \\[5pt]
\frac{\partial w_{\bar{\alpha}}}{\partial y^\beta}=0 \\[5pt]
\frac{\partial w_{\bar{\alpha}}}{\partial z^a}=0
\end{array}\right.
\end{equation}
These systems admit a solution if and only if the following integrability conditions (with $\beta$ different from $\alpha$ and $\bar{\alpha}$) respectively hold:
\begin{eqnarray*}
\frac{\D R_a}{\D z^b}=\frac{\D R_a}{\D x^\alpha}=\frac{\D R_a}{\D y^\alpha}&=&0\,,\\
\frac{\D R_\alpha}{\D z^a}=\frac{\D R_\alpha}{\D x^\beta}=\frac{\D R_\alpha}{\D y^\beta}=
\frac{\D R_\alpha}{\D z^{\bar{\alpha}}}&=&0\,,\\
\frac{\D R_{\bar{\alpha}}}{\D z^a}=\frac{\D R_{\bar{\alpha}}}{\D x^\beta}=\frac{\D R_{\bar{\alpha}}}{\D y^\beta}=\frac{\D R_{\bar{\alpha}}}{\D x^\alpha}&=&0\,.
\end{eqnarray*}
Due to the linearity of the total derivative with respect to $z^\mu$, the necessary and sufficient conditions for the existence of a solution of systems (\ref{sistema}) become:
$$
\frac{\D R_\mu}{\D z^\nu}=\frac{\partial R_\mu}{\partial z^\nu}+R_\nu\frac{\partial R_\mu}{\partial w_\nu}=0
$$
where no summation over the repeated indices is understood. Substituting the expression of $R_\mu$ from (\ref{sistema}) the Levi--Civita condition is obtained. 
\end{Dim}
\end{Teo}
\begin{Prop}
If the Hamiltonian $H$ is real and satisfies the Levi--Civita equations then the complete integral can be chosen to be real.
\begin{Dim}
The definition (\ref{sistema}) of $R_\mu$ proves that if $H$ is real then $R_a$ are real and $\overline{R}_\alpha=R_{\bar{\alpha}}$. Hence, the first of the systems (\ref{sis_reale}) implies that $w_a$ can always be chosen real. The case of $w_\alpha$, instead, is more subtle: first one must observe that in the separated integral $W$ the function $W_\alpha$ always appears added to the function $W_{\bar{\alpha}}$, then one must consider the function $w_\alpha-\overline{w}_{\bar{\alpha}}$ that has the same imaginary part of $w_\alpha+w_{\bar{\alpha}}$. The last two of the systems (\ref{sis_reale}) imply that
\begin{eqnarray*}
\frac{\partial}{\partial x^\alpha}(w_\alpha-\overline{w}_{\bar{\alpha}})&=&R_\alpha-\overline{R}_{\bar{\alpha}}=0\,; \\
\frac{\partial}{\partial y^\alpha}(w_\alpha-\overline{w}_{\bar{\alpha}})&=&iR_\alpha-i\overline{R}_{\bar{\alpha}}=0\,.
\end{eqnarray*}
Hence the imaginary part of $w_\alpha+w_{\bar{\alpha}}=W^\prime_\alpha+W^\prime_{\bar{\alpha}}$ is a constant that can be chosen to be zero and this means that also the imaginary part of $W_\alpha+W^{\bar{\alpha}}$ is a constant and it can be chosen to be zero. It follows that always exist a real complete integral of the Hamilton--Jacobi equation.
\end{Dim}
\end{Prop}
\begin{Oss}
The previous theorem holds for any system of canonical variables $(z^\mu,P_\mu)$ independently from their association with eigenforms of a Killing tensor and from the signature of the metric $\T{g}$, provided the non real variables are pairwise conjugate.
\end{Oss}
\begin{Teo}
The functions $K_{(\sigma)}=\frac{1}{2}\sum_\mu\lambda_{(\sigma)}^\mu g^\mu{P_\mu}^2$ e $H=\frac{1}{2}\sum_\nu g^\nu {P_\nu}^2$ are in involution if and only if the Eisenhart conditions hold:
\begin{eqnarray}
&&g^\nu\frac{\partial\lambda_{(\sigma)}^\nu}{\partial z^\mu}=(\lambda_{(\sigma)}^\mu-\lambda_{(\sigma)}^\nu)\frac{\partial g^\nu}{\partial z^\mu}\label{Eisenhart1}\\
\iff&&\frac{\partial\lambda_{(\sigma)}^\nu}{\partial z^\mu}=(\lambda_{(\sigma)}^\mu-\lambda_{(\sigma)}^\nu)\frac{\partial}{\partial z^\mu}\ln |g^\nu|\label{Eisenhart2}\\
\iff&&\frac{\partial}{\partial z^\mu}(g^\nu\lambda_{(\sigma)}^\nu)=\lambda_{(\sigma)}^\mu\frac{\partial g^\nu}{\partial z^\mu}\,.\label{Eisenhart3}
\end{eqnarray}
\begin{Dim}
Using the Poisson bracket properties one obtains
\begin{eqnarray*}
\{K_{(\sigma)},H\} &=& \frac{1}{2}\textstyle{\sum_\nu}\{K_{(\sigma)},g^\nu\}{P_\nu}^2+
\textstyle{\sum_\nu}\{K_{(\sigma)},P_\nu\}P_\nu g^\nu \\
&=&\frac{1}{4}\textstyle{\sum_{\mu\nu}}\{\lambda_{(\sigma)}^\mu g^\mu,g^\nu\}{P_\mu}^2{P_\nu}^2+
\displaystyle{\frac{1}{2}}\textstyle{\sum_{\mu\nu}}\{P_\mu,g^\nu\}\lambda^\mu_{(\sigma)}g^\mu P_\mu{P_\nu}^2 +\\
&&\frac{1}{2}\textstyle{\sum_{\mu\nu}}\{\lambda_{(\sigma)}^\mu g^\mu,P_\nu\}g^\nu P_\nu{P_\mu}^2 +
\displaystyle{\frac{1}{2}}\textstyle{\sum_{\mu\nu}}\{{P_\mu}^2,P_\nu\}\lambda_{(\sigma)}^\mu g^\mu g^\nu P_\nu \\
&=&\frac{1}{2}\textstyle{\sum_{\mu\nu}}\left[
(\lambda_{(\sigma)}^\mu-\lambda_{(\sigma)}^\nu)\frac{\partial g^\nu}{\partial z^\mu}-g^\nu\frac{\partial\lambda_{(\sigma)}^\nu}{\partial z^\mu}
\right]g^\mu P_\mu{P_\nu}^2
\end{eqnarray*}
and the thesis follows from the identity of polynomials.
\end{Dim}
\end{Teo}

The existence of a KS-algebra implies the equations (\ref{Eisenhart1}-\ref{Eisenhart3}). The converse rises some monodromy problem in the definition of the functions $\lambda_{(\sigma)}^\nu$ that will not be examined here. In the following, the existence of a KS-algebra or of a characteristic Killing tensor will be understood, together with the associated complex variables.

By writing the Hamilton--Jacobi equations for all the Hamiltonians $K_{(1)}=H,\ldots,K_{(n)}$ associated to a KS-algebra the following system of PDEs is obtained
\BE\label{Systemsep}
\left(\begin{array}{ccc}
g^1 & \cdots & g^n \\
\lambda_{(2)}^1g^1 & \cdots & \lambda_{(2)}^ng^n \\
\vdots && \vdots \\
\lambda_{(n)}^1g^1 & \cdots & \lambda_{(n)}^ng^n
\end{array}\right)
\left(\begin{array}{c}
(\partial_1W)^2\\ (\partial_2W)^2\\ \vdots\\ (\partial_nW)^2
\end{array}\right)
=
\left(\begin{array}{c}
c_{1}\\ c_{2}\\ \vdots\\ c_{n}
\end{array}\right).
\EE
Denoting by $\T{S}$  the inverse of the matrix in the left side of the equation, one has
\BE\label{RastelliDim}
\left\{\begin{array}{ll}
\sum_\nu S^\tau_\nu g^\nu=\delta^\tau_1 & \sigma=1 \\
\sum_\nu S^\tau_\nu \lambda_{(\sigma)}^\nu g^\nu=\delta^\tau_\sigma & \sigma\neq1
\end{array}\right.
\EE
\begin{Teo}
The Eisenhart equations (\ref{Eisenhart3}) hold if and only if the entries of the invertible matrix $\T{S}$ satisfy $\frac{\partial}{\partial z^\mu} S^\tau_\nu=0$ for $\mu\neq\nu$.
\begin{Dim}
The differentiation of the two equations (\ref{RastelliDim}) gives, respectively
$$
\left\{\begin{array}{ll}
\sum_\nu g^\nu\frac{\partial}{\partial z^\mu} S^\tau_\nu+\sum_\nu S^\tau_\nu \frac{\partial}{\partial z^\mu} g^\nu=0 & \sigma=1 \\
\sum_\nu  \lambda_{(\sigma)}^\nu g^\nu\frac{\partial}{\partial z^\mu} S^\tau_\nu+
\sum_\nu S^\tau_\nu\frac{\partial}{\partial z^\mu} (\lambda_{(\sigma)}^\nu g^\nu)=0
& \sigma\neq1
\end{array}\right.
$$
The second expression can be written as
$$
\textstyle{\sum_\nu} S^\tau_\nu\frac{\partial}{\partial z^\mu} (\lambda_{(\sigma)}^\nu g^\nu)
-\textstyle{\sum_\nu} S^\tau_\nu \lambda_{(\sigma)}^\mu\frac{\partial}{\partial z^\mu} g^\nu 
+\textstyle{\sum_\nu} S^\tau_\nu \lambda_{(\sigma)}^\mu\frac{\partial}{\partial z^\mu}  g^\nu
+\textstyle{\sum_\nu} \lambda_{(\sigma)}^\nu g^\nu \frac{\partial}{\partial z^\mu} S^\tau_\nu =0
$$
and subtracting the first expression (multiplied by $\lambda_{(\sigma)}^\mu$) one has, with $\sigma\neq1$
$$
\textstyle{\sum_\nu} \Big[\frac{\partial}{\partial z^\mu} (\lambda_{(\sigma)}^\nu g^\nu)
- \lambda_{(\sigma)}^\mu\frac{\partial}{\partial z^\mu} g^\nu\Big] S^\tau_\nu
+\textstyle{\sum_\nu}  \Big(\lambda_{(\sigma)}^\nu- \lambda_{(\sigma)}^\mu\Big) g^\nu\frac{\partial}{\partial z^\mu}S^\tau_\nu
=0\,.
$$
Therefore, if $\partial_\mu S^\tau_\nu=0$ for $\mu\neq \nu$ the second summation vanishes and the invertibility of $\T{S}$ imply the Eisenhart equations.

On the converse, taking the Eisenhart conditions as hypotheses for any fixed value of $\mu$ and $\tau$ one has the system of $n-1$ equations
$$
\sum_{\nu\neq \mu}\left[\T{M}^{(\mu)}\right]^\nu_\sigma u_\nu^{(\mu \tau)}=0,\quad \sigma\neq1
$$
in the $n-1$ unknown $u_\nu^{(\mu\,\tau)}=\partial_\mu S^\tau_\nu$ with $\nu\neq \mu$. The determinant of the matrix
\begin{eqnarray*}
&&\T{M}^{(\mu)}=\Big((\lambda^\nu_{(\sigma)}-\lambda^\mu_{(\sigma)})g^\nu\Big)=\\
&&
\hspace{-10pt}\left(\begin{array}{cccccc}
g^1\\[-5pt]
&\hspace{-10pt}\ddots\\[-5pt]
&&\hspace{-10pt}g^{\mu-1}\\[-5pt]
&&&\hspace{-15pt}g^{\mu+1}\\[-5pt]
&&&&\hspace{-15pt}\ddots\\[-5pt]
&&&&&\hspace{-10pt}g^n
\end{array}\right)
\left(\begin{array}{cccccc}
\lambda_{(2)}^1\!\!-\!\!\lambda_{(2)}^\mu&\!\!\!\cdots\!\!\!&
\lambda_{(2)}^{\mu-1}\!\!-\!\!\lambda_{(2)}^\mu\!\!&\!\!\lambda_{(2)}^{\mu+1}\!\!-\!\!\lambda_{(2)}^\mu
&\!\!\!\cdots\!\!\!&\lambda_{(2)}^n\!\!-\!\!\lambda_{(2)}^\mu\\
\vdots&&\vdots&\vdots&&\vdots\\
\vdots&&\vdots&\vdots&&\vdots\\
\lambda_{(n)}^1\!\!-\!\!\lambda_{(n)}^\mu&\!\!\!\cdots\!\!\!&
\lambda_{(n)}^{\mu-1}\!\!-\!\!\lambda_{(n)}^\mu\!\!&\!\!\lambda_{(n)}^{\mu+1}\!\!-\!\!\lambda_{(n)}^\mu
&\!\!\!\cdots\!\!\!&\lambda_{(n)}^n\!\!-\!\!\lambda_{(n)}^\mu
\end{array}\right)
\end{eqnarray*}
is
\begin{eqnarray*}
&&\left|\begin{array}{cccccc}
\lambda_{(2)}^1\!\!-\!\!\lambda_{(2)}^\mu&\!\!\!\cdots\!\!\!&
\lambda_{(2)}^{\mu-1}\!\!-\!\!\lambda_{(2)}^\mu\!\!&\!\!\lambda_{(2)}^{\mu+1}\!\!-\!\!\lambda_{(2)}^\mu
&\!\!\!\cdots\!\!\!&\lambda_{(2)}^n\!\!-\!\!\lambda_{(2)}^\mu\\
\vdots&&\vdots&\vdots&&\vdots\\
\vdots&&\vdots&\vdots&&\vdots\\
\lambda_{(n)}^1\!\!-\!\!\lambda_{(n)}^\mu&\!\!\!\cdots\!\!\!&
\lambda_{(n)}^{\mu-1}\!\!-\!\!\lambda_{(n)}^\mu\!\!&\!\!\lambda_{(n)}^{\mu+1}\!\!-\!\!\lambda_{(n)}^\mu
&\!\!\!\cdots\!\!\!&\lambda_{(n)}^n\!\!-\!\!\lambda_{(n)}^\mu
\end{array}\right|\;\prod_{\nu\neq \mu}g^\nu\\
&&=\left|\begin{array}{ccccccc}
0 && 0&1&0 && 0 \\
\lambda_{(2)}^1\!\!-\!\!\lambda_{(2)}^\mu&\!\!\!\cdots\!\!\!&
\lambda_{(2)}^{\mu-1}\!\!-\!\!\lambda_{(2)}^\mu\!\!&
\lambda_{(2)}^\mu
&\!\!\lambda_{(2)}^{\mu+1}\!\!-\!\!\lambda_{(2)}^\mu
&\!\!\!\cdots\!\!\!&\lambda_{(2)}^n\!\!-\!\!\lambda_{(2)}^\mu\\
\vdots&&\vdots&\vdots&\vdots&&\vdots\\
\vdots&&\vdots&\vdots&\vdots&&\vdots\\
\lambda_{(n)}^1\!\!-\!\!\lambda_{(n)}^\mu&\!\!\!\cdots\!\!\!&
\lambda_{(n)}^{\mu+1}\!\!-\!\!\lambda_{(n)}^\mu\!\!&
\lambda_{(n)}^\mu
&\!\!\lambda_{(n)}^{\mu+1}\!\!-\!\!\lambda_{(n)}^\mu
&\!\!\!\cdots\!\!\!&\lambda_{(n)}^n\!\!-\!\!\lambda_{(n)}^\mu
\end{array}\right|\;\prod_{\nu\neq \mu}g^\nu\\
&&=\left|\begin{array}{ccccc}
1 && 1 && 1 \\
\lambda_{(2)}^1&\!\!\!\cdots\!\!\!&
\lambda_{(2)}^\mu
&\!\!\!\cdots\!\!\!&\lambda_{(2)}^n\\
\vdots&&\vdots&&\vdots\\
\vdots&&\vdots&&\vdots\\
\lambda_{(n)}^1&\!\!\!\cdots\!\!\!&
\lambda_{(n)}^\mu
&\!\!\!\cdots\!\!\!&\lambda_{(n)}^n
\end{array}\right|\;\prod_{\nu\neq\mu}g^\nu
\end{eqnarray*}
and it vanishes if and only if the determinant of $\T{S}$ vanishes. The unique solution of the system (\ref{Systemsep}) is then $u^{(\mu\,\tau)}_\nu=0$ for all $\mu, \tau, \nu$. 
\end{Dim}
\end{Teo}

The matrix $\T{S}$, where the $\mu$-th row depends only from $z^\mu$, is the complex generalization of the classical St\"ackel matrix. Taking the inverse of (\ref{Systemsep}) the separated ODE's
$$
\displaystyle \frac{\partial W}{\partial z^\nu}=\frac {dW_\nu}{dz^\nu}=S^\tau _\nu c_\tau ,
$$
are found. Because the function $W$ is real it is possible to apply the classical Jacobi method to the function $W(z^\mu(q^\sigma),c_\nu)$ and solve the Hamilton equations. The complex variables are introduced only in order to find a separated complete integral of the Hamilton--Jacobi equation. Since this integral is found, being it a real function, everything can be done in real coordinates and a complex version of the Jacobi theorem is not needed.

\section{Examples}
\subsection{Two-dimensional case}

Because complex variables with non-vanishing imaginary part always appear as conjugate pairs, the two dimensional case is particularly relevant and is a model for each pair of truly complex variables.

Let us consider the particularly simple case of a two-dimensional manifold $Q$ on which the function $z^1=z$ has everywhere non-vanishing imaginary part. Then the relation $z^2=\bar{z}$ necessarily holds and the coordinates $(x,y)$ are defined by
\begin{eqnarray*}
z &=& x+iy \\
\bar{z} &=& x-iy
\end{eqnarray*}
The matrix $\T{J}$, in this case is
$$
\big(J^\nu_\mu\big)=\frac{i}{2J}
\left(\begin{array}{cc}
\frac{\partial \bar{z}}{\partial q^2} &-\frac{\partial \bar{z}}{\partial q^1} \\
-\frac{\partial z}{\partial q^2} & \frac{\partial z}{\partial q^1}
\end{array}\right)
$$
where
$$
J(q^\mu) =\frac{\partial x}{\partial q^1}\frac{\partial y}{\partial q^2}-\frac{\partial x}{\partial q^2}\frac{\partial y}{\partial q^1}=\frac{i}{2}
\left|\begin{array}{cc}
\frac{\partial z}{\partial q^1} & \frac{\partial \bar{z}}{\partial q^1} \\[5pt]
\frac{\partial z}{\partial q^2} & \frac{\partial \bar{z}}{\partial q^2}
\end{array}\right|.
$$
The vector field $\frac{\partial}{\partial z}$ is given by
$$
\frac{\partial}{\partial z}=\frac{i}{2J}\left(\frac{\partial \bar{z}}{\partial q^2}\frac{\partial}{\partial q^1}-\frac{\partial \bar{z}}{\partial q^1}\frac{\partial}{\partial q^2}\right)
$$
and the vector fields $\frac{\partial}{\partial x}$ e $\frac{\partial}{\partial y}$ can be written as
\begin{eqnarray*}
\frac{\partial}{\partial x} &=& \frac{1}{J}\left(\frac{\partial y}{\partial q^2}\frac{\partial}{\partial q^1}-\frac{\partial y}{\partial q^1}\frac{\partial}{\partial q^2}\right) \\
\frac{\partial}{\partial y} &=& \frac{1}{J}\left(\frac{\partial x}{\partial q^1}\frac{\partial}{\partial q^2}-\frac{\partial x}{\partial q^2}\frac{\partial}{\partial q^1}\right) \\
\end{eqnarray*}

The link between old and new momenta is given by
$$
p_\mu=\frac{\partial z}{\partial q^\mu}P+\frac{\partial \bar{z}}{\partial q^\mu}\bar{P}
$$
that imply
\BE\label{momento_inv}
P=\frac{i}{2J}\left(\frac{\partial \bar{z}}{\partial q^2}p_1-\frac{\partial \bar{z}}{\partial q^1}p_2\right)\,.
\EE

The Poisson bracket of $P$ with $z$ and $\bar{z}$ can be calculated directly from the formula (\ref{momento_inv}) and the definition of $J$:
\begin{eqnarray*}
\{P,z\} &=& \frac{\partial P}{\partial p_\mu}\frac{\partial z}{\partial q^\mu}-
\frac{\partial P}{\partial q^\mu}\frac{\partial z}{\partial p_\mu}\\
&=& \frac{i}{2J}\left(\frac{\partial \bar{z}}{\partial q^2}\frac{\partial z}{\partial q^1}-
\frac{\partial \bar{z}}{\partial q^1}\frac{\partial z}{\partial q^2}\right)\\
&=&\frac{i}{2J}\frac{2J}{i}=1\\
\{P,\bar{z}\} &=& \frac{\partial P}{\partial p_\mu}\frac{\partial \bar{z}}{\partial q^\mu}-
\frac{\partial P}{\partial q^\mu}\frac{\partial \bar{z}}{\partial p_\mu} \\
&=&\frac{i}{2J}\left(\frac{\partial \bar{z}}{\partial q^2}\frac{\partial \bar{z}}{\partial q^1}-
\frac{\partial \bar{z}}{\partial q^1}\frac{\partial \bar{z}}{\partial q^2}\right)\\
&=&0
\end{eqnarray*}
as expected for a canonical transformation.

Finally the Hamiltonians $H$ and $K$ associated to the metric and the Killing tensor are, in the $x,y$ coordinates:
\begin{eqnarray*}
H&=&\frac{1}{2}g^{\mu\nu}\left(\frac{\partial x}{\partial q^\mu}p_x+\frac{\partial y}{\partial q^\mu}p_y\right)
\left(\frac{\partial x}{\partial q^\nu}p_x+\frac{\partial y}{\partial q^\nu}p_y\right) \\
&=&\frac{1}{2}g^{\mu\nu}\frac{\partial x}{\partial q^\mu}\frac{\partial x}{\partial q^\nu}p_x^2+
g^{\mu\nu}\frac{\partial x}{\partial q^\mu}\frac{\partial y}{\partial q^\nu}p_xp_y+
\frac{1}{2}g^{\mu\nu}\frac{\partial y}{\partial q^\mu}\frac{\partial y}{\partial q^\nu}p_y^2
\end{eqnarray*}
but because $x=\frac{1}{2}(z+\bar{z})$ e $y=\frac{i}{2}(\bar{z}-z)$ one has
\begin{eqnarray*}
4g^{\mu\nu}\frac{\partial x}{\partial q^\mu}\frac{\partial x}{\partial q^\nu}&=&
g^{\mu\nu}\left(\frac{\partial z}{\partial q^\mu}+\frac{\partial \bar{z}}{\partial q^\mu}\right)
\left(\frac{\partial z}{\partial q^\nu}+\frac{\partial \bar{z}}{\partial q^\nu}\right)=g+\bar{g}\\
4g^{\mu\nu}\frac{\partial x}{\partial q^\mu}\frac{\partial y}{\partial q^\nu}&=&
ig^{\mu\nu}\left(\frac{\partial z}{\partial q^\mu}+\frac{\partial \bar{z}}{\partial q^\mu}\right)
\left(\frac{\partial \bar{z}}{\partial q^\nu}-\frac{\partial z}{\partial q^\nu}\right)=i(\bar{g}-g) \\
4g^{\mu\nu}\frac{\partial y}{\partial q^\mu}\frac{\partial y}{\partial q^\nu}&=&
-g^{\mu\nu}\left(\frac{\partial \bar{z}}{\partial q^\mu}-\frac{\partial z}{\partial q^\mu}\right)
\left(\frac{\partial \bar{z}}{\partial q^\nu}-\frac{\partial z}{\partial q^\nu}\right)=-g-\bar{g}
\end{eqnarray*}
that is
\begin{eqnarray*}
H&=&\frac{1}{8}(g+\bar{g})p_x^2+\frac{i}{4}(\bar{g}-g)p_xp_y-\frac{1}{8}(g+\bar{g})p_y^2=\frac{1}{2}\Big(gP^2+\bar{g}\overline{P}^2\Big)\\
K&=&\frac{1}{8}(\lambda g+\bar{\lambda}\bar{g})p_x^2+\frac{i}{4}(\bar{\lambda}\bar{g}-\lambda g)p_xp_y-\frac{1}{8}(\lambda g+\bar{\lambda}\bar{g})p_y^2=
\frac{1}{2}\Big(\lambda gP^2+\bar{\lambda}\bar{g}\overline{P}^2\Big)
\end{eqnarray*}
Hence, the coordinates $x$ and $y$ (differently from the variables $z$ and $\bar{z}$) do not diagonalize neither the metric nor the Killing tensor.

A complete separated integral of the Hamilton--Jacobi equation associated to $H$ is a function $W(z,\bar{z};h,k)$ such that, if $c_1=h$ and $c_2=k$,
\begin{eqnarray*}
&&W(z,\bar{z};h,k)=W_1(z;h,k)+W_2(\bar{z};h,k) \\[5pt]
&&H\left(P_z=\frac{\partial W}{\partial z},\bar{P}_z=\frac{\partial W}{\partial \bar{z}}\right)=h\\[5pt]
&&\det \left(\frac{\partial^2 W}{\partial z^\mu\partial c_\nu}\right)\neq0
\end{eqnarray*}
The fact that $\T{K}$ is a Killing tensor for the metric $\T{g}$ is equivalent to the involutivity between the functions $H$ and $K$, hence the equivalent Eisenhart conditions take the form
\BE\label{Eisenhart}
\frac{\partial \lambda_\nu}{\partial z^\mu}=(\lambda_\mu-\lambda_\nu)\frac{\partial}{\partial z^\mu}\ln |g^\nu|
\EE
where no summation over repeated indices is assumed. The equations (\ref{Eisenhart}) correspond to the two (complex conjugate) systems
$$
\left\{\begin{array}{l}
\displaystyle{\frac{\partial \lambda}{\partial z}=0} \\[10pt]
\displaystyle{g\frac{\partial \lambda}{\partial\bar{z}}=(\bar{\lambda}-\lambda)\frac{\partial g}{\partial\bar{z}}}
\end{array}\right.
\qquad
\left\{\begin{array}{l}
\displaystyle{\frac{\partial\bar{\lambda}}{\partial\bar{z}}=0} \\[10pt]
\displaystyle{\bar{g}\frac{\partial\bar{\lambda}}{\partial z}=(\lambda-\bar{\lambda})\frac{\partial \bar{g}}{\partial z}}
\end{array}\right.
$$
or, equivalently, to the two systems
\BE\label{sis_Eisenhart}
\left\{\begin{array}{l}
\displaystyle{\frac{\partial}{\partial z}\,\lambda=0} \\[10pt]
\displaystyle{\frac{\partial}{\partial z}\Big[\bar{g}(\lambda-\bar{\lambda})\Big]=0}
\end{array}\right.
\qquad
\left\{\begin{array}{l}
\displaystyle{\frac{\partial}{\partial\bar{z}}\;\bar{\lambda}=0} \\[10pt]
\displaystyle{\frac{\partial}{\partial\bar{z}}\Big[g(\lambda-\bar{\lambda})\Big]=0}
\end{array}\right.
\EE
The two Hamilton--Jacobi equations for the two Hamiltonians $H$ and $K$, in the variables $z$ and $\bar{z}$, give the PDE system
$$
\left\{\begin{array}{l}
\displaystyle{g\left(\frac{\partial W}{\partial z}\right)^2+\bar{g}\left(\frac{\partial W}{\partial\bar{z}}\right)^2=h} \\[10pt]
\displaystyle{\lambda g\left(\frac{\partial W}{\partial z}\right)^2+\bar{\lambda}\bar{g}\left(\frac{\partial W}{\partial\bar{z}}\right)^2=k}
\end{array}\right.
$$
that is by solving with respect to the derivatives
$$
\left(\begin{array}{c}
\left(\frac{\partial W}{\partial z}\right)^2\\ \left(\frac{\partial W}{\partial\bar{z}}\right)^2
\end{array}\right)=
\left(\begin{array}{cc}
\frac{\bar{\lambda}}{g(\bar{\lambda}-\lambda)} & \frac{1}{g(\lambda-\bar{\lambda})} \\
\frac{\lambda}{\bar{g}(\lambda-\bar{\lambda})} & \frac{1}{\bar{g}(\bar{\lambda}-\lambda)}
\end{array}\right)
\left(\begin{array}{c}
h\\ k
\end{array}\right)
$$
The systems (\ref{sis_Eisenhart}) imply that the first row of the matrix depends only from $z$, while the second one only from $\bar{z}$, hence if $W_1(z)$ and $W_2(\bar{z})$ satisfies
\begin{eqnarray*}
\left(\frac{\partial W_1}{\partial z}\right)^2&=&\frac{h\bar{\lambda}-k}{g(\bar{\lambda}-\lambda)}\\
\left(\frac{\partial W_2}{\partial\bar{z}}\right)^2&=&\frac{h\lambda-k}{\bar{g}(\lambda-\bar{\lambda})}
\end{eqnarray*}
the function $W(z,\bar{z})=W_1(z)+W_2(\bar{z})$ is a real solution of the Hamilton--Jacobi equation for both the Hamiltonians $H$ and $K$.

It is important to observe that, being $\frac{\partial W_2}{\partial\bar{z}}$ the complex conjugate of $\frac{\partial\bar{W_2}}{\partial z}$ and being $h$ and $k$ real-valued, the function $\bar{W_2}$ satisfies the same equation of $W_1(z)$ and so the solution of the Hamilton--Jacobi equation is found integrating just one equation. Again appears the phenomenon of redundancy.

Moreover, if $\lambda$ is not constant from (\ref{sis_Eisenhart}) follows that $\lambda$ is a function of $\bar{z}$ only. By taking $\lambda=\bar{z}$ it is possible to determinate a standard form for $g$:
$$
g=\frac{A(z)}{z-\bar{z}}\,
$$
where $A$ is an arbitrary function of $z$ and, consequently, olomorphic.

\subsection{First example: everywhere complex variables}
Let consider, in the Minkowski plane with coordinates $q^1=x$ and $q^2=t$, the metric and the Killing tensor
$$
\big(g^{\mu\nu}\big)=\left(\begin{array}{cc}
1&0\\ 0&-1
\end{array}\right)
\quad
\big(K^{\mu\nu}\big)=\left(\begin{array}{cc}
0&1\\ 1&0
\end{array}\right)
$$
The corresponding Hamiltonian functions are
$$
H=\frac{1}{2}(p_x^2-p_t^2) \quad K=p_xp_t
$$
The equation $\T{K}\,\D z^\mu=\lambda^\mu\T{g}\,\D z^\mu$ relates to the eigenvalues $\pm i$ the functions $z^1=z=x+it$ and $z^2=\bar{z}=x-it$. The eigenvectors of $\T{K}$ have non-vanishing imaginary part everywhere the change of variables is invertible, thus no real separable coordinates can be found.
 
The Jacobian matrix and its inverse $\T{J}$ are
$$
\left(\frac{\partial z^\nu}{\partial q^\mu}\right)=\left(
\begin{array}{cc}
1 & i \\
1 & -i
\end{array}
\right)\,,\qquad
\big(J^\mu_\nu\big)=\frac{1}{2}\left(
\begin{array}{cc}
1 & 1 \\
-i & i
\end{array}
\right)
$$
then, the momenta $p_x$ and $p_t$ are given by
$$
p_x=P+\bar{P} \quad p_t=i(P-\bar{P})\,.
$$
In the new variables the Hamiltonians become
$$
H=P^2+\bar{P}^2 \quad K=i(P^2-\bar{P}^2)\,.
$$
The two Hamilton--Jacobi equations are
$$
\left\{\begin{array}{l}
\left(\dfrac{\partial W}{\partial z}\right)^2+\left(\dfrac{\partial W}{\partial\bar{z}}\right)^2=h \\[10pt]
\left(\dfrac{\partial W}{\partial z}\right)^2-\left(\dfrac{\partial W}{\partial\bar{z}}\right)^2=-ik
\end{array}\right.
$$
and looking for a separated solution $W=W_1(z)+W_2(\bar{z})$ one has that $W_1$ and $\overline{W}_2$ (taking $h$ and $k$ real) both satisfy the equation
$$
\left(\frac{\D S}{\D z}\right)^2=\frac{h-ik}{2}\,.
$$
Set $\zeta=h+ik$ (not belonging to the negative real semi-axis, to ensure the monodromy of the square root) the solution of the previous equation is
$$
S=z\sqrt{\frac{\zeta}{2}}
$$
that is (see \S 21.2 in \cite{DNF}): 
\begin{eqnarray*}
W=\frac{1}{\sqrt{2}}\left(z\sqrt{\zeta}+\bar{z}\sqrt{\bar{\zeta}}\right)&=&
\sqrt{2}\,\mathrm{Re}\left(z\sqrt{\zeta}\right)\\
&=&\sqrt{2}\left(
x\,\mathrm{Re}\sqrt{\zeta}-t\,\mathrm{Im}\sqrt{\zeta}\right)\\
&=&x\sqrt{h+\sqrt{h^2+k^2}}+\frac{kt}{\sqrt{h+\sqrt{h^2+k^2}}}
\end{eqnarray*}

To find the conjugate momenta for $h$ and $k$ one can apply the Jacobi theorem, being $W$, $h$ and $k$ real:
\begin{eqnarray*}
\theta_h=\frac{\partial W}{\partial h}&=&
\frac{1}{|\zeta|\sqrt{2}}\mathrm{Re}\left(z\overline{\sqrt{\zeta}}\right)\\
&=& \frac{1}{|\zeta|\sqrt{2}}\left(x\,\mathrm{Re}\sqrt{\zeta}+t\,\mathrm{Im}\sqrt{\zeta}\right)\\
\theta_k=\frac{\partial W}{\partial k}
&=&\frac{1}{|\zeta|\sqrt{2}}\mathrm{Im}\left(z\overline{\sqrt{\zeta}}\right)\\
&=& \frac{1}{|\zeta|\sqrt{2}}\left(t\,\mathrm{Re}\sqrt{\zeta}-x\,\mathrm{Im}\sqrt{\zeta}\right)
\end{eqnarray*}
where $\theta_k$ is constant while $\theta_h=\tau-\tau_0$.

By solving with respect to $x$ and $t$ one has
\begin{eqnarray*}
x &=& \sqrt{2}\left(\theta_h\mathrm{Im}\sqrt{\zeta}+\theta_k\mathrm{Re}\sqrt{\zeta}\right)\\
t &=& \sqrt{2}\left(\theta_h\mathrm{Re}\sqrt{\zeta}-\theta_k\mathrm{Im}\sqrt{\zeta}\right)
\end{eqnarray*}
and the integral curves are (as expected) straight lines.

\subsection{Second example: parabolic coordinates}
Let us consider, in the Minkowski plane with coordinates $q^1=x$ and $q^2=t$, the metric and the Killing tensor
$$
\big(g^{\mu\nu}\big)=\left(\begin{array}{cc}
1&0\\ 0&-1
\end{array}\right)
\quad
\big(K^{\mu\nu}\big)=\left(\begin{array}{cc}
1+2t & x+t \\
x+t &1+2x
\end{array}\right)
$$
The corresponding Hamiltonian functions are
$$
H=\frac{1}{2}(p_x^2-p_t^2) \quad K=\frac{1}{2}(p_x^2+p_t^2)+(p_x+p_t)(xp_t+tp_x)\,.
$$
The functions $z_1, z_2$ are found by solving the equation
\mbox{$\T{K}\,\D z^\mu=\lambda^\mu \T{g}\,\D z^\mu$},
that relates to the eigenvalues $\lambda^\mu=t-x-(-1)^\mu\sqrt{1+2(x+t)}$ the functions $z^\mu=x-t-(-1)^\mu\sqrt{1+2(x+t)}$. It is worth to observe that $\lambda^1$ depends only from $z^2$ e vice versa. Moreover the two eigenvectors are  real and distinct or complex conjugate depending on the sign of $2(x+t)+1$; the line $x+t=-\frac{1}{2}$ consists of singular points where the eigenvectors coincide. Then, when $\lambda^\mu$ are real, real separable coordinates are well-defined, the proposed approach allows to use the same notation also when the variables are complex conjugate.

The Jacobian matrix and its inverse $\T{J}$ are
\begin{eqnarray*}
\left(\frac{\partial z^\nu}{\partial q^\mu}\right)&=&\left(
\begin{array}{cc}
1+\dfrac{1}{\sqrt{1+2(x+t)}} & -1+\dfrac{1}{\sqrt{1+2(x+t)}} \\
1-\dfrac{1}{\sqrt{1+2(x+t)}} & -1-\dfrac{1}{\sqrt{1+2(x+t)}}
\end{array}
\right)\\
\big(J^\mu_\nu\big)&=&\frac{1}{4}\left(
\begin{array}{cc}
1+\sqrt{1+2(x+t)} & 1-\sqrt{1+2(x+t)} \\
-1+\sqrt{1+2(x+t)} & -1-\sqrt{1+2(x+t)}
\end{array}
\right)
\end{eqnarray*}
hence, the momenta $p_x$ and $p_t$ are given by
$$
p_x=P_1+P_2+2\,\frac{P_1-P_2}{z^1-z^2} \quad p_t=-P_1-P_2+2\,\frac{P_1-P_2}{z^1-z^2}
$$
and the Hamiltonians by
$$
H=4\,\frac{{P_1}^2-{P_2}^2}{z^1-z^2}\,, \quad K=4\,\frac{z^2{P_1}^2-z^1{P_2}^2}{z^2-z^1}\,.
$$
The two Hamilton--Jacobi equations become
$$
\left\{\begin{array}{l}
\left(\dfrac{\partial W}{\partial z^1}\right)^2-\left(\dfrac{\partial W}{\partial z^2}\right)^2=\dfrac{1}{4}h(z^1-z^2) \\
z^2\left(\dfrac{\partial W}{\partial z^1}\right)^2-z^1\left(\dfrac{\partial W}{\partial z^2}\right)^2=\dfrac{1}{4}k(z^2-z^1)
\end{array}\right.
$$
and by solving them one has (outside the singular line $2(x+t)+1=0$) the two equations
$$
\left(\frac{\partial W}{\partial z^\mu}\right)^2=\frac{z^\mu h-k}{4}
$$
that admit a separated solution $W=W_1(z^1)+W_2(z^2)$. The two functions $W_\mu$ both satisfy the equation 
$$
\left(\frac{\D S}{\D z}\right)^2=\frac{z h-k}{4}
$$
that is, by introducing the variable $\zeta=zh-k$ (assumed not belonging to the real negative semi-axis for monodromy reasons), the equation
$$
\frac{\D S}{\D\zeta}=\Phi(\zeta)=\frac{1}{2h}\sqrt{\zeta}\,.
$$
The solution of this equation is
$$
S(\zeta)=\frac{1}{3h}\sqrt{\zeta^3}=\frac{1}{3h}\sqrt{{(zh-k)}^3}
$$
then, if $\zeta^\mu=z^\mu h-k$ are both real one has
$$
W=S(\zeta^1)+S(\zeta^2)=\frac{1}{3h}\sqrt{{(z^1h-k)}^3}+\frac{1}{3h}\sqrt{{(z^2h-k)}^3}\,,
$$
while if they are complex conjugate one obtains
$$
W=\frac{1}{3h}\mathrm{Re}(\zeta^1)^{\frac{3}{2}}\,.
$$
However, it is not strictly necessary to solve the differential equation in order to apply the Jacobi method:
\begin{eqnarray*}
\theta_h=\frac{\partial W}{\partial h}&=&
\frac{\partial W}{\partial \zeta^1}\frac{\partial \zeta^1}{\partial h}+\frac{\partial W}{\partial \zeta^2}\frac{\partial \zeta^2}{\partial h}=
\Phi(\zeta^1)\zeta^1+\Phi(\zeta^2)\zeta^2\\
\theta_k=\frac{\partial W}{\partial k}&=&
\frac{\partial W}{\partial \zeta^1}\frac{\partial \zeta^1}{\partial k}+\frac{\partial W}{\partial \zeta^2}\frac{\partial \zeta^2}{\partial k}=
-\Phi(\zeta^1)-\Phi(\zeta^2)
\end{eqnarray*}
where $\theta_k$ is constant while $\theta_h=\tau-\tau_0$.

After substituting the values of $\zeta^\mu$ as functions of $x$ and $t$ and solving with respect to these two variables one obtains the straight lines
\begin{eqnarray*}
x&=& -\frac{1+2h\theta_k^2}{6h\theta_k}\,\theta_h-\frac{h\theta_k^2-1}{3}h\theta_k^2+\frac{k}{2h}-\frac{1}{4}\\
t&=& \frac{1-2h\theta_k^2}{6h\theta_k}\,\theta_h-\frac{h\theta_k^2+1}{3}h\theta_k^2-\frac{k}{2h}-\frac{1}{4}
\end{eqnarray*}

Finally, with the choice $h=-k=\dfrac{1}{2{\theta_k}^2}$ one obtains the solution
$$
t=0\,,\quad x=-\frac{2}{3}(\theta_k\theta_h+1)
$$
that is a geodesic defined both in the region where $z^\mu$ are complex-valued and in the region where they are real-valued.

\section{Web associated to complex variables}
In this section the level curve of real and imaginary part of complex variables associated to Killing tensor in the Minkowski plane are plotted and compared with the webs corresponding to real eigenvalues. The names of coordinate systems are taken from \cite{CDM}.

\begin{figure}[!h]
\hspace{1cm}
\includegraphics[width=4.5cm]{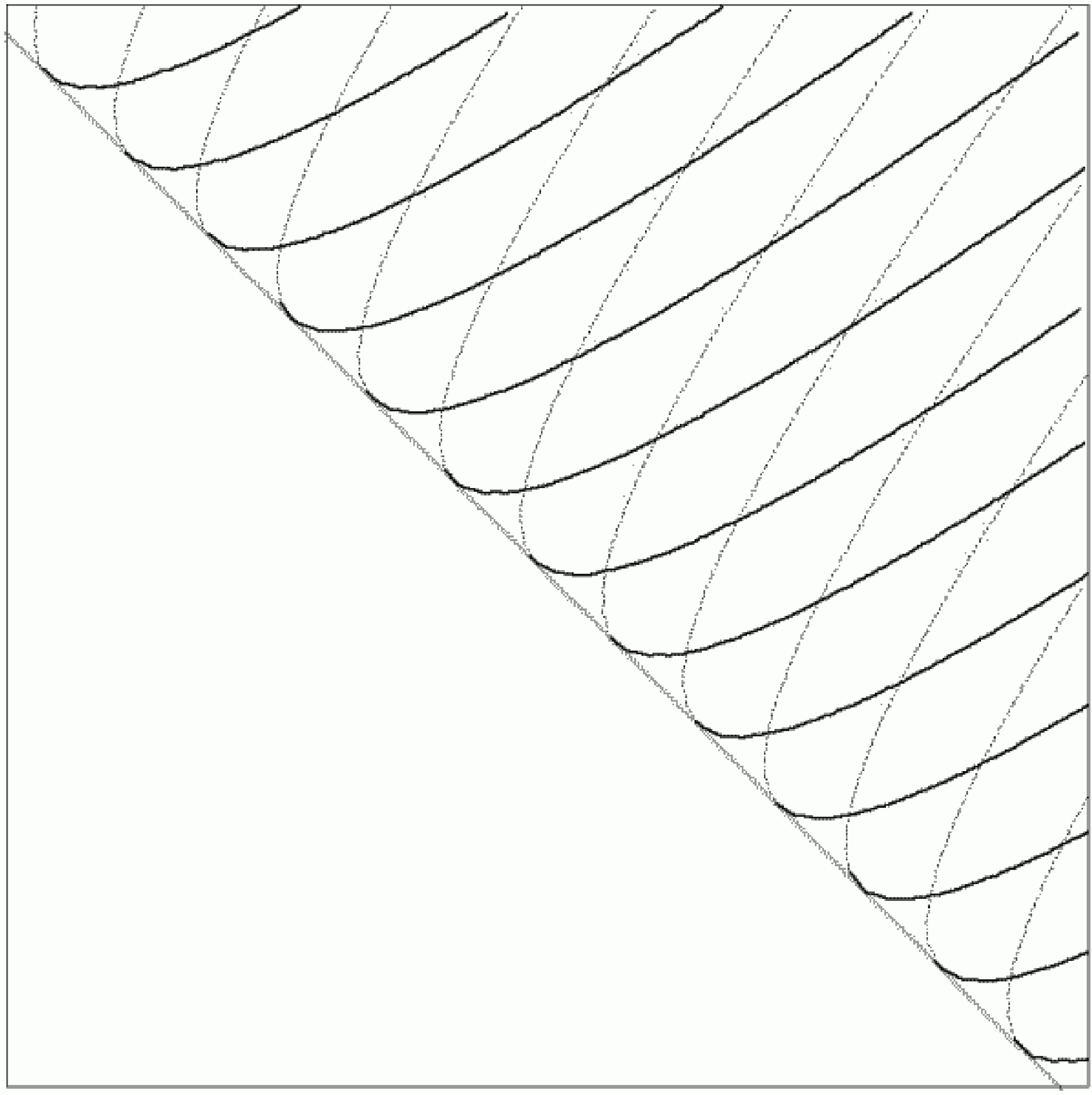}
\hspace{0.7cm}
\includegraphics[width=4.5cm]{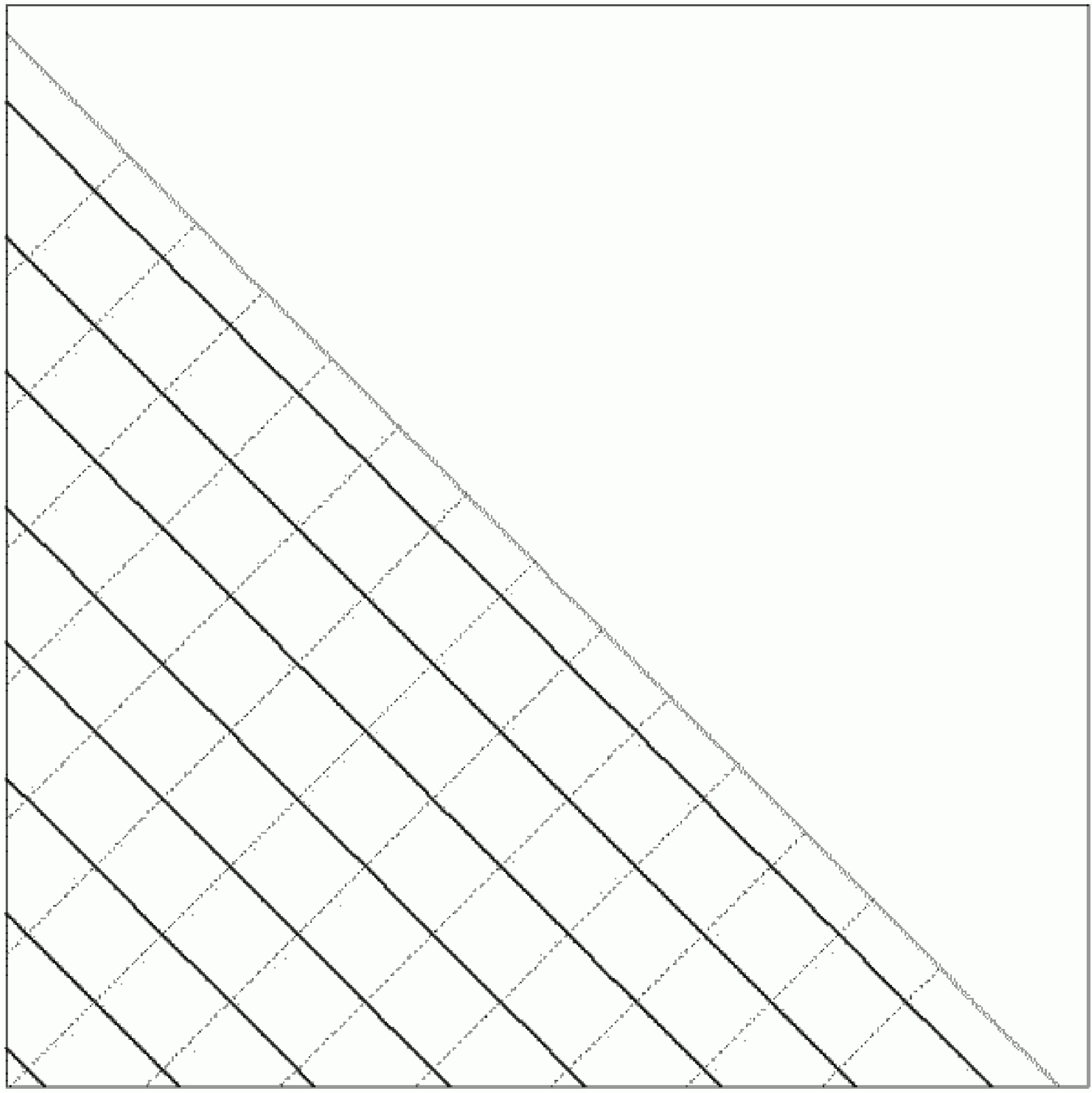}
\caption{Real and complex webs for SC3}
\end{figure}
\begin{figure}[!h]
\hspace{1cm}
\includegraphics[width=4.5cm]{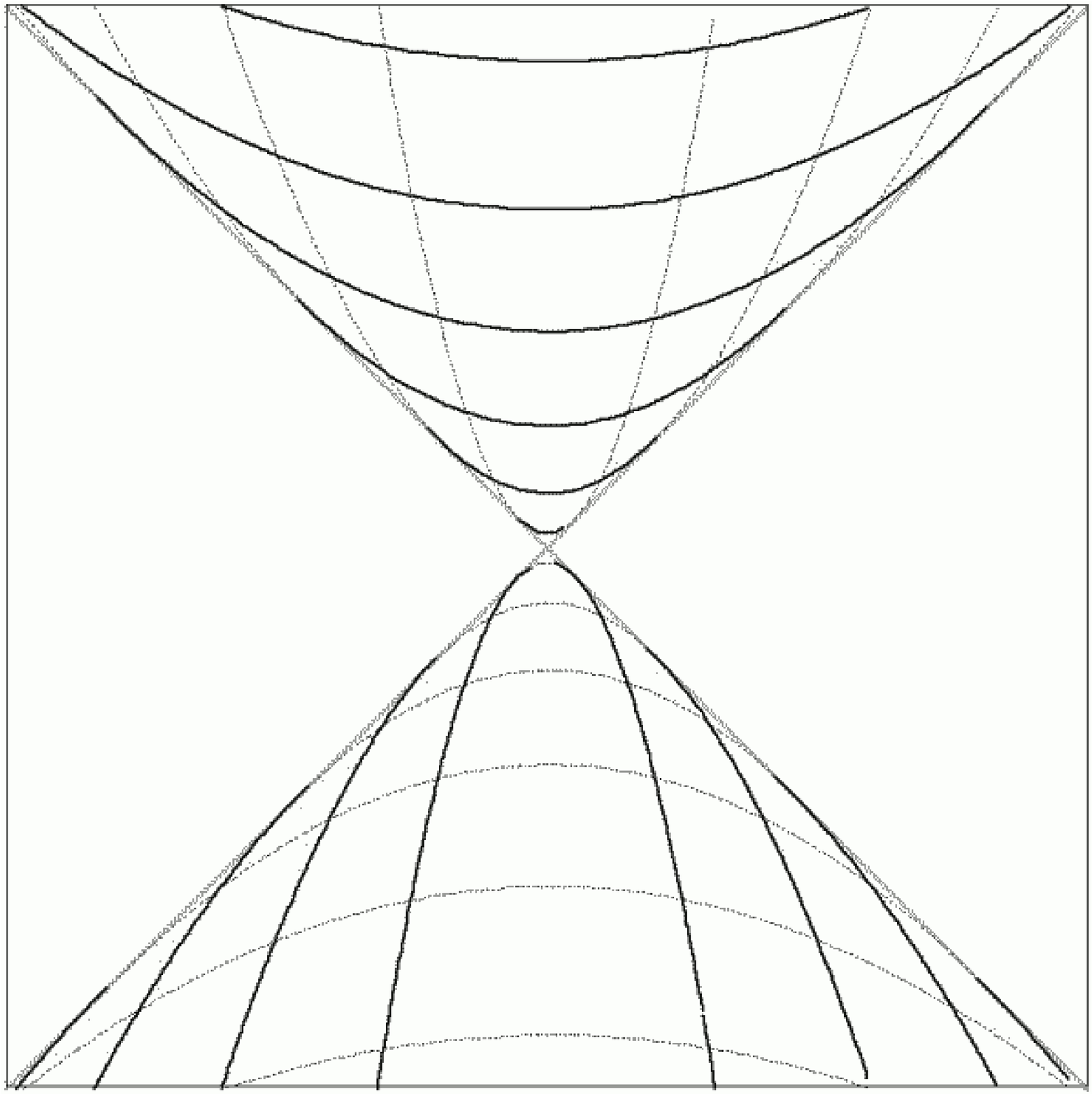}
\hspace{0.7cm}
\includegraphics[width=4.5cm]{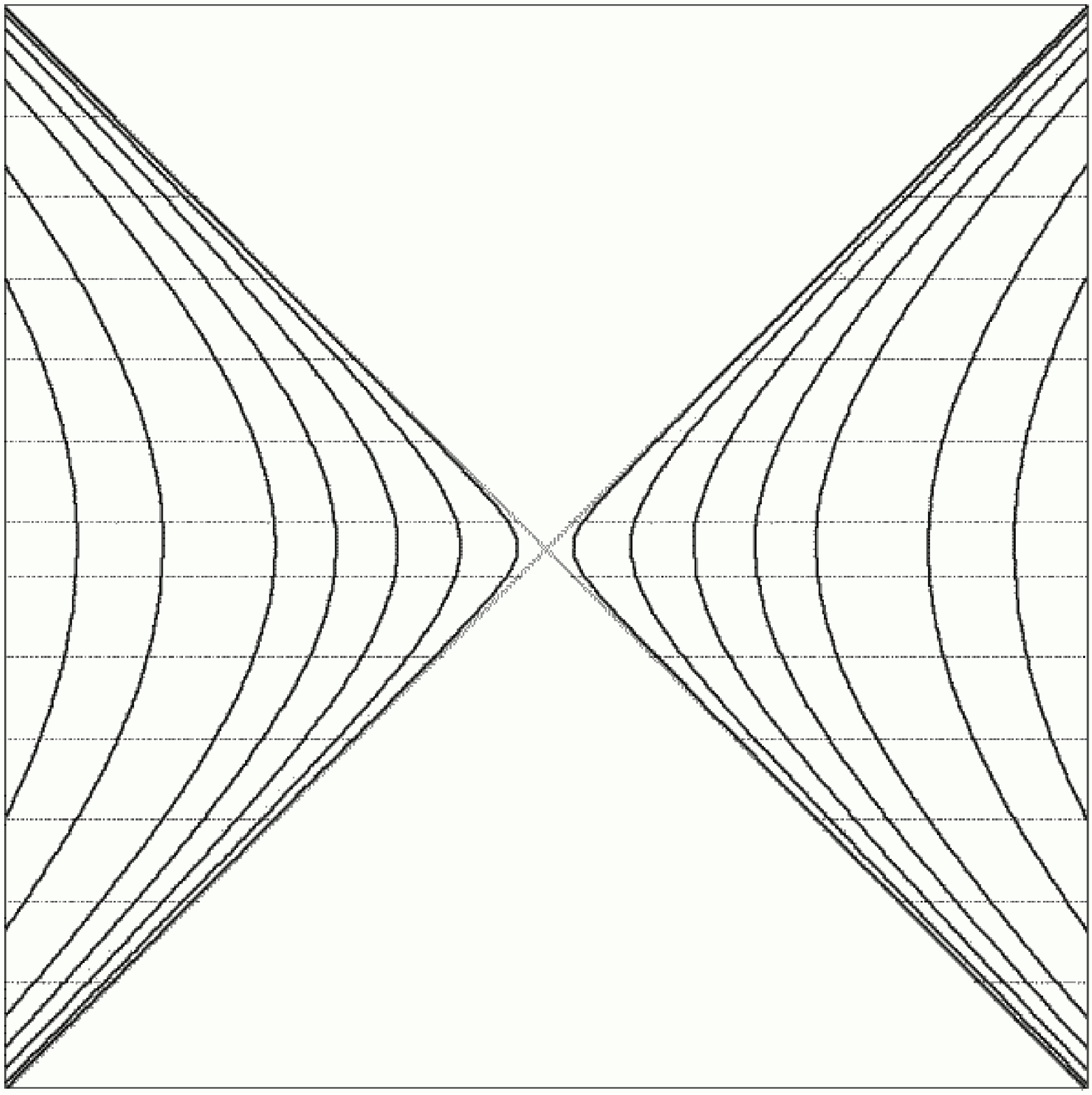}
\caption{Real and complex webs for SC4}
\end{figure}
\begin{figure}[!h]
\hspace{1cm}
\includegraphics[width=4.5cm]{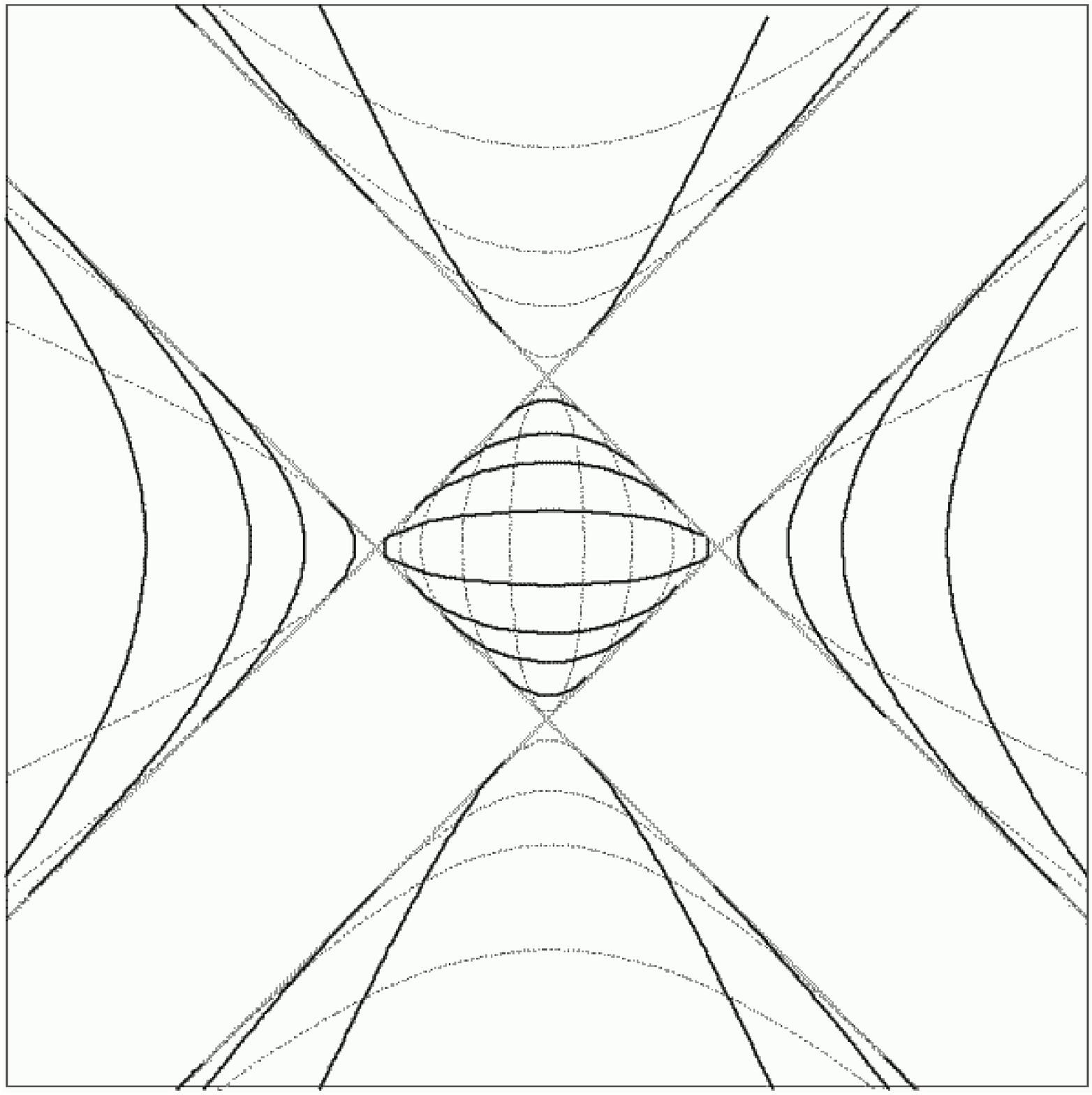}
\hspace{0.7cm}
\includegraphics[width=4.5cm]{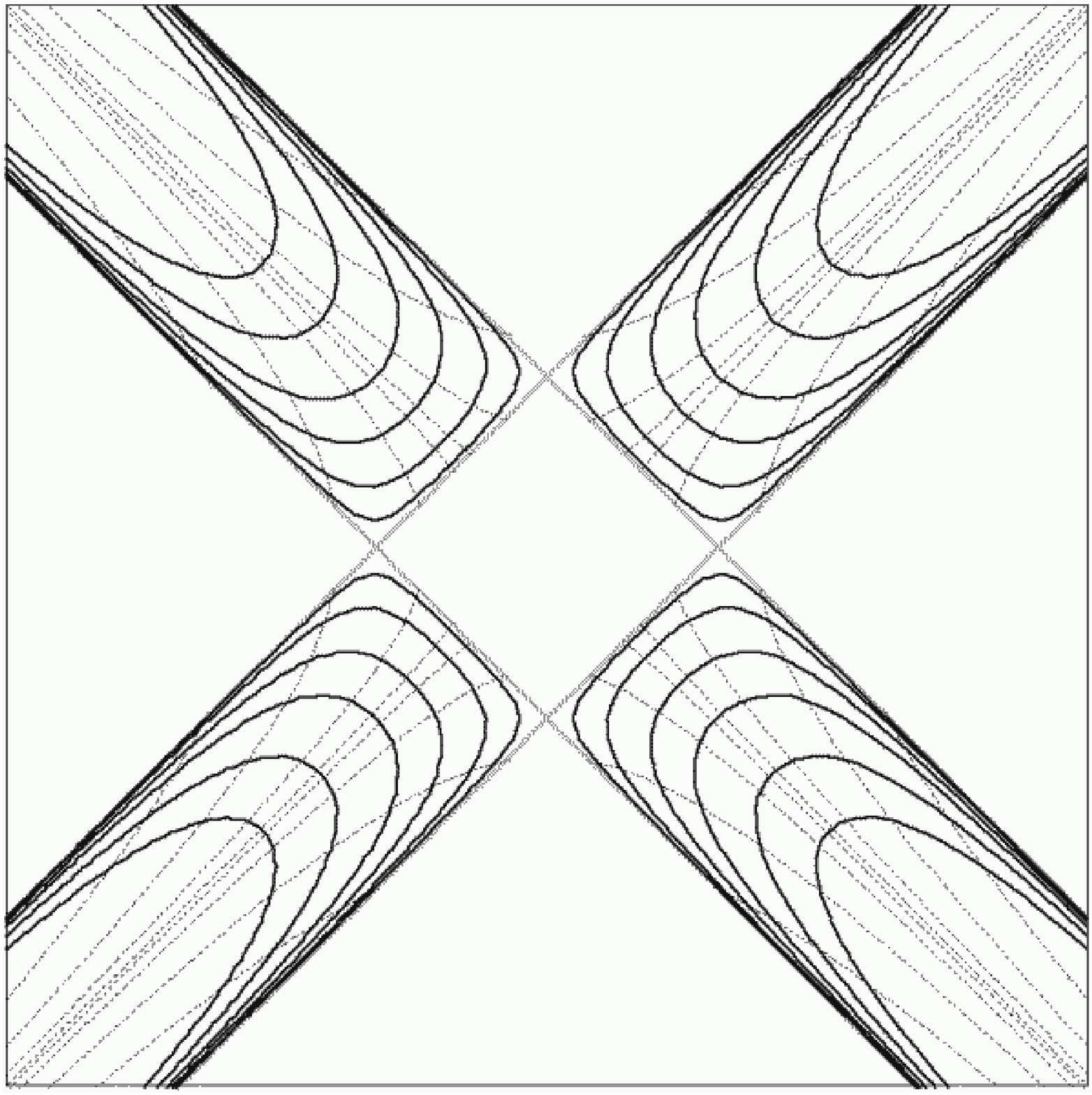}
\caption{Real and complex webs for SC5 and 10}
\end{figure}
\begin{figure}[!h]
\hspace{1cm}
\includegraphics[width=4.5cm]{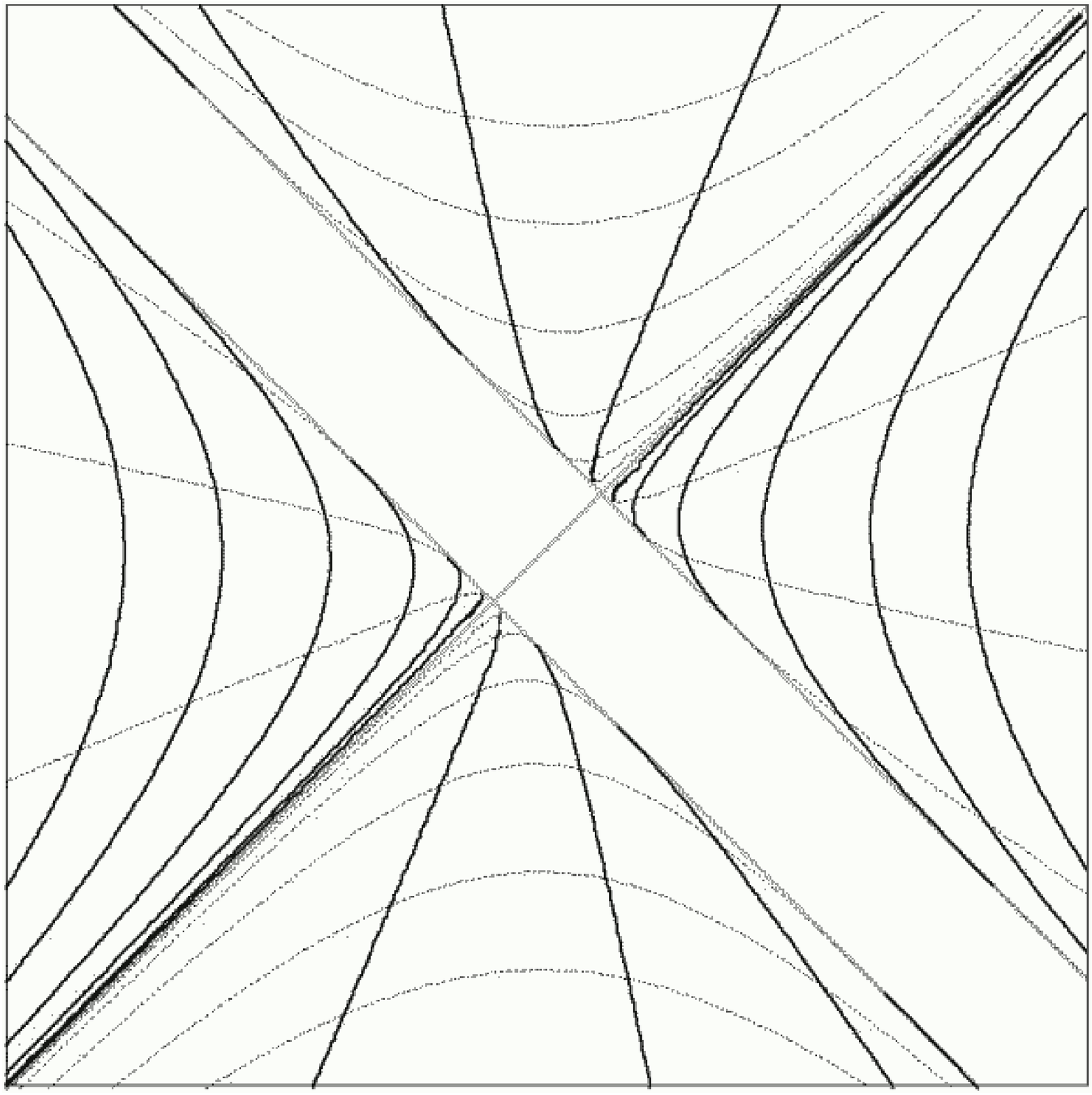}
\hspace{0.7cm}
\includegraphics[width=4.5cm]{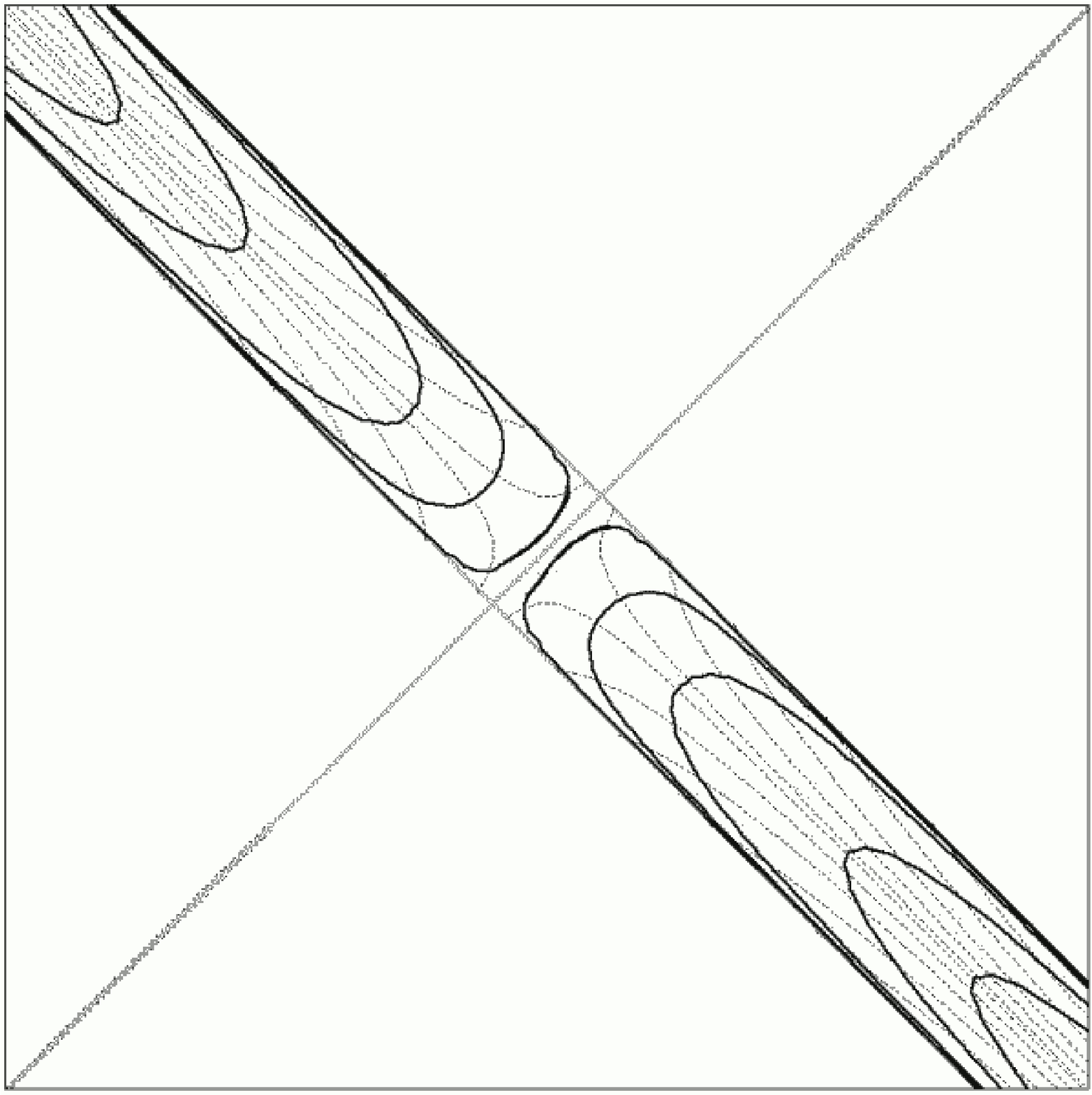}
\caption{Real and complex webs for SC7}
\end{figure}
\begin{figure}[!h]
\hspace{1cm}
\includegraphics[width=4.5cm]{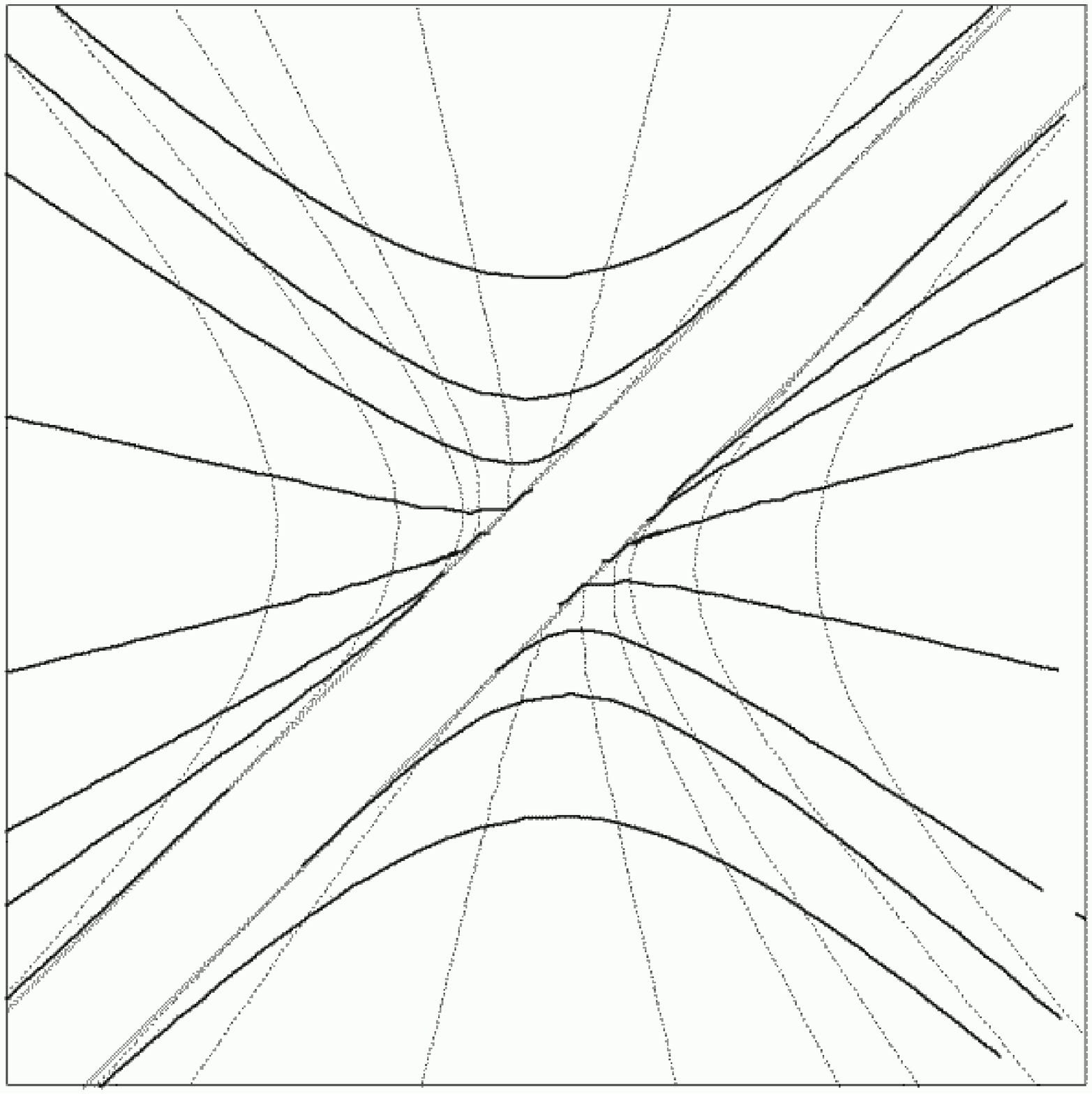}
\hspace{0.7cm}
\includegraphics[width=4.5cm]{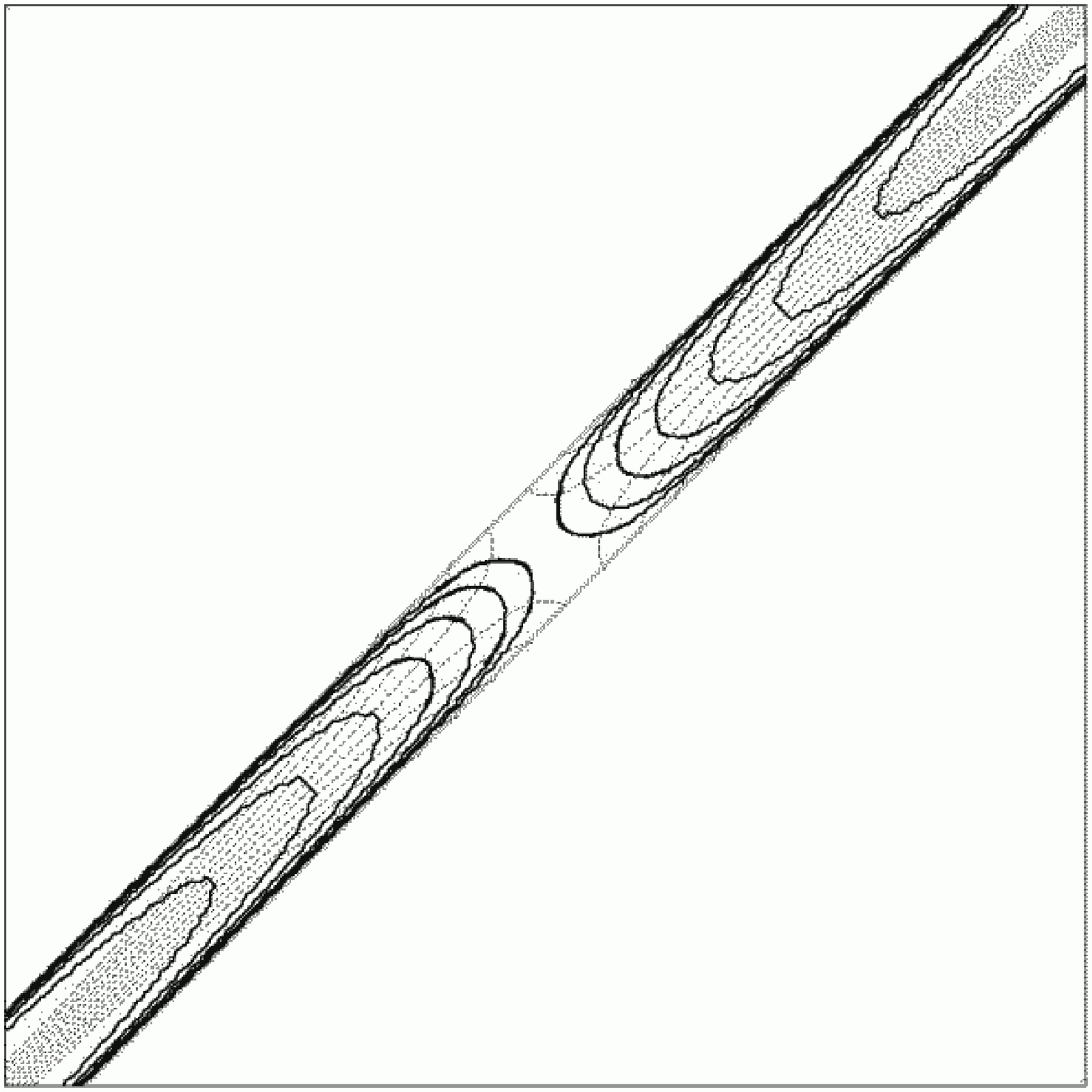}
\caption{Real and complex webs for SC8}
\end{figure}

\pagebreak
\section{Final remarks}
In this paper a method is presented to obtain, using a set of complex variables, a complete solution of the Hamilton--Jacobi equation on a pseudo-Riemannian manifold. The complex variables arise from Killing tensors with pointwise distinct eigenvalues and normal eigenvectors, giving a more general definition of characteristic Killing tensor, starting with a real Hamiltonan is always possible to obtain a real complete integral, separated in the complex variables but not in their real and imaginary parts. The method can be extended to Hamiltonians with scalar and vector potentials and to the study of Schr\"odinger equation through the techniques of \cite{BCR1} and \cite{BCR2}.

\section*{Acknowledgments}
The authors wish to thank dr. Claudia Chanu, prof. Sergio Benenti, prof. Stefan Rauch-Wojciechowski and prof. Domenico Delbosco for useful discussions, remarks and suggestions.

\end{document}